\documentclass[
]{elsarticle}
\usepackage[
reviewcopy
]{adndt}

\usepackage{longtable}

\usepackage[utf8]{inputenc}
\usepackage[T1]{fontenc}
\usepackage{microtype}

\usepackage{xcolor}
\usepackage[colorlinks=true,linkcolor=blue,urlcolor=blue,citecolor=blue,bookmarks=true,bookmarksopenlevel=2,breaklinks=true]{hyperref}

\usepackage{mathptmx}

\usepackage{amsmath}
\usepackage{amssymb}
\usepackage{array,booktabs,tabularx}

\biboptions{square,sort&compress}
\bibpunct[]{[}{]}{,}{n}{}{;}
\usepackage{xspace}
\newcommand{\grasp}{\textsc{Grasp2018}\xspace}
\newcommand{\fac}{\textsc{Fac}\xspace}

\newcommand{\Wviii}{W~VIII\xspace}
\newcommand{\Wix}{W~IX\xspace}
\newcommand{\Wx}{W~X\xspace}

\newcommand{\Wxiii}{W~XIII\xspace}
\newcommand{\cmm}{cm\textsuperscript{-1}\xspace}

\usepackage{siunitx}

\usepackage{mhchem}

\definecolor{lightblue}{rgb}{.90,.95,1}
\usepackage{framed} 
\colorlet{shadecolor}{lightblue}

\usepackage{soul} 
\sethlcolor{lightblue}
\setstcolor{red}

\begin{document}

\begin{frontmatter}

\journal{Atomic Data and Nuclear Data Tables}


\title{Theoretical level energies, radiative lifetimes and transitions in \Wix}

  \author{Karol Kozio{\l}\corref{cor}}
  \ead{E-mail: karol.koziol@ncbj.gov.pl}

  \cortext[cor]{Corresponding author.}

  \author{Jacek Rzadkiewicz}

  \address{Narodowe Centrum Bada\'{n} J\k{a}drowych (NCBJ), Andrzeja So{\l}tana 7, 05-400 Otwock-\'{S}wierk, Poland}

\date{2020} 

\begin{abstract}  
The atomic states of the \Wix (W\textsuperscript{8+}) tungsten ion lying below the W\textsuperscript{9+} ionisation threshold have been studied theoretically, employing the multiconfiguration Dirac--Hartree--Fock method with configuration interaction. 
The level electronic structures and their energies are presented. 
The electric dipole (E1), magnetic dipole (M1), electric quadrupole (E2), and magnetic quadrupole (M2) radiative transitions have been computed in order to calculate the radiative lifetimes of given states. 
Transition wavelengths, energies, and decay rates are also presented for selected high-intensity E1 transitions. 
The configuration interaction method was applied to estimate electron correlation effects. 
The aim of the present research was to fill a lack of atomic data for low-charged tungsten ions, which may be useful in low-temperature plasma diagnostics and may form the base for collisional-radiative modelling of spectra for low-charged tungsten ions. 
\end{abstract}

\end{frontmatter}

\clearpage

\tableofcontents
\listofDtables
\listofDfigures
\vskip5pc


\section{Introduction}

Tungsten, with its high melting point, high thermal resistance, low erosion rate, and low tritium retention, has been chosen as the plasma-facing material for the ITER divertor \cite{Matthews2007,Bolt2002,Ralchenko2013}. However, due to its high radiation power the tungsten concentration in the core fusion plasma should not exceed $10^{-4}$, to control radiation losses \cite{Putterich2008}. Despite the low erosion rate, tungsten ions will appear in the plasma. In the ITER divertor region a relatively cool plasma with large temperature gradients from a few eV up to 150~eV is expected \cite{Divertor1999,Clementson2010c}. Under these conditions, emerging tungsten ions with charge states up to $q=10+$ can be a valuable source of information about key parameters of divertor plasma. Therefore, the atomic data pertaining to low-charged tungsten ions is of growing interest in tokamak plasma diagnostics \cite{Clementson2010c,Clementson2015,Lu2019}. 
Low-charged tungsten ions from \Wviii (W\textsuperscript{7+}\xspace) to \Wxiii (W\textsuperscript{12+}\xspace) are of interest because their outermost orbitals 4f and 5p are close to each other, and configurations \ce{[Kr] 4d^10 4f^14 5s^2 5p^x} and \ce{[Kr] 4d^10 4f^13 5s^2 5p^{x+1}} compete for the ground state, as was reported first by Sugar and Kaufman \cite{Sugar1975}. Moreover, visible transitions in high-$Z$ ions, for example, low-charged tungsten ions, are suggested as potential candidates for a precise atomic clock that can be used in laboratory searches for the time variation of the fine-structure constant \cite{Berengut2011,Safronova2018}.

In recent years, a few studies were focused on experimental and theoretical research on \Wviii ions. 
Berengut et al. \cite{Berengut2011} calculated the energies for four levels with configurations \ce{4f^14 5s^2 5p^5} and \ce{4f^13 5s^2 5p^6} and confirmed the level sequence reported by Kramida and Shirai \cite{Kramida2009}. 
Afterwards, Ryabtsev et al. \cite{Ryabtsev2013,Ryabtsev2015} observed the spectrum of W\textsuperscript{7+} and reported the appropriate level energies. 
Then, the \Wviii level structure was analysed theoretically by Deprince and Quinet \cite{Deprince2015} by using the pseudo-relativistic Hartree--Fock method. 
The EUV spectrum of \Wviii was measured by Clementson et al. \cite{Clementson2015} and studied theoretically by using a modified version of the Cowan code \cite{Cowan1981}.  
Mita et al. \cite{Mita2017} measured the fine-structure splitting of the ground term \ce{4f^13 5p^6 ^2F_{5/2,7/2}} in \Wviii. 
The fine-structure splitting was also measured by Lu et al. \cite{Lu2019} and reproduced by the multiconfiguration Dirac--Hartree--Fock (MCDHF) calculations with configuration interaction (CI), with good agreement. 
In that work, the evolution of the low-charge states of tungsten ions occurring in low-temperature plasmas is also discussed. The given charge-state fraction may be enriched as a result of indirect ionisation caused by stepwise excitation between intermediate metastable states of the lower charge state of W ions. 

In recent years, the complex structure of \Wix ion levels was also studied experimentally and theoretically. 
Berengut et al. \cite{Berengut2011} calculated the energies for 16 levels belonging to configurations \ce{4f^14 5s^2 5p^4}, \ce{4f^13 5s^2 5p^5}, and \ce{4f^12 5s^2 5p^6}. 
Ryabtsev et al. \cite{Ryabtsev2015} reported many spectral lines of the \Wix ion in the range of 170--200~\AA. In order to analyse these lines, they performed calculations by the Hartree--Fock method with relativistic corrections implemented in the Cowan code \cite{Cowan1981} and concluded that the observed lines may be assigned to \ce{4f^{14\ldots11} 5s^2 5p^{3\ldots6} 5d} $\to$ \ce{4f^{14\ldots12} 5s^2 5p^{3\ldots6}} transitions types. 
Afterwards, Mita et al. \cite{Mita2017} measured some of the \Wix lines in EUV and visible ranges, however, without assigning these lines to the appropriate atomic levels. 
Nonetheless, there is still a lack of tungsten spectroscopic data for some low-ionisation stages, as pointed out by Kramida and Shirai \cite{Kramida2009} and by Ralchenko \cite{Ralchenko2013}. 
In the Atomic Spectra Database of the National Institute of Standards and Technology \cite{NIST_ASD}, only data for the ground states are available for \Wix to \Wxiii spectra.

Here we present the MCDHF-CI calculations of the energy levels in the \Wix (W\textsuperscript{8+}\xspace) tungsten ion. 
The electric dipole (E1), magnetic dipole (M1), electric quadrupole (E2), and magnetic quadrupole (M2) radiative transitions have been computed in order to calculate the radiative lifetimes of states. 
In more detail, we analysed selected high-intensity E1 transitions, for which transition wavelengths, energies and decay rates are presented. 
The aim of the present research was to fill a lack of atomic data for low-charged tungsten ions, which may be useful in low-temperature plasma diagnostics and may form the base for collisional-radiative modelling of spectra for low-charged tungsten ions.

\begin{table*}[!htb]
\centering
\caption{Valence electron configurations studied in present work.}\label{tab:conf}
\begin{tabular*}{\linewidth}{@{}l@{\extracolsep{\fill}} ll ll@{}}
\toprule
Block & \multicolumn{2}{c}{Even states} & \multicolumn{2}{c}{Odd states} \\
\cmidrule{2-3}\cmidrule{4-5}
& Configurations & Number of levels & Configurations & Number of levels \\
\midrule
Block 1 & \ce{4f^{14\ldots12} 5s^2 5p^{4\ldots6}} & 30 & &  \\
Block 2 &  && \ce{4f^{14\ldots13} 5s^1 5p^{5\ldots6}} & 8 \\
Block 3 &  && \ce{4f^{14\ldots11} 5s^{2} 5p^{3\ldots6} 5d^{1}} & 1328 \\
Block 4 & \ce{4f^{14\ldots11} 5s^{2} 5p^{3\ldots6} 5f^{1}} & 1704 & \ce{4f^{14\ldots12} 5s^{1} 5p^{4\ldots6} 5f^{1}} & 622 \\
& \ce{4f^{14\ldots12} 5s^{1} 5p^{4\ldots6} 5d^{1}} & 496 & \ce{4f^{14\ldots11} 5s^{2} 5p^{3\ldots6} 6s^{1}} & 290 \\
& \ce{4f^{14\ldots11} 5s^{2} 5p^{3\ldots6} 6p^{1}} & 846 &  &  \\
& \ce{4f^{14\ldots10} 5s^{2} 5p^{2\ldots6} 5d^{2}} & 21925 &  &  \\
& \ce{4f^{14} 5p^{6}} & 1 &  &  \\
\bottomrule
\end{tabular*}
\end{table*}

\setlength{\LTcapwidth}{\linewidth}
\setlength{\LTleft}{0pt}
\setlength{\LTright}{0pt} 
\setlength{\tabcolsep}{0.5\tabcolsep}
\renewcommand{\arraystretch}{1.0}

\begin{longtable}{@{\extracolsep{\fill}} lll l *{4}{r} @{\extracolsep{3pt}} *{2}{r} @{\extracolsep{3pt}} *{2}{r} @{}}
\caption{Energy of lowest lying atomic levels of \Wix obtained from different theoretical approaches and from the experiment.\label{tab:Wix-conv}}
&&&& \multicolumn{4}{c}{\grasp code} & \multicolumn{2}{c}{\fac code} & \multicolumn{2}{c}{other} \\
\cmidrule{5-8}\cmidrule{9-10}\cmidrule{11-12}
No.  &  $J$  &  $p$  &  Term  &  EOL MR  &  EAL MR  & EOL CI &  EAL CI  &  EAL MR & EOL MR &  Ref. \cite{Berengut2011}  &  Exp. \cite{Nakamura_priv} \\
\midrule
\endfirsthead
\multicolumn{12}{l}{Table \ref{tab:Wix-conv} (continued)}\\
\midrule
&&&& \multicolumn{4}{c}{\grasp code} & \multicolumn{2}{c}{\fac code} & \multicolumn{2}{c}{other} \\
\cmidrule{5-8}\cmidrule{9-10}\cmidrule{11-12}
No.  &  $J$  &  $p$  &  Term  &  EOL MR  &  EAL MR  & EOL CI &  EAL CI  &  EAL MR & EOL MR &  Ref. \cite{Berengut2011}  &  Exp. \cite{Nakamura_priv} \\
\midrule
\endhead
\midrule
\endfoot
\bottomrule
\endlastfoot
%
%
\multicolumn{12}{l}{Levels (\cmm)}\\
1 & 2 &  +  &   \ce{4f^14 5s^2 5p^4 ^3P}   & 0 & 39652 & 0 & 0 & 0 & 0 & 0 &  \\
2 & 4 &  +  &   \ce{4f^13 5s^2 5p^5 ^3F}   & 9440 & 0 & 20310 & 11159 & 6165 & 31470 & 6075 &  \\
3 & 3 &  +  &   \ce{4f^13 5s^2 5p^5 ^1F}   & 10445 & 1059 & 20682 & 11578 & 6994 & 32309 & 6357 &  \\
4 & 5 &  +  &   \ce{4f^13 5s^2 5p^5 ^3G}   & 14766 & 5511 & 25168 & 16073 & 10982 & 36298 & 11122 &  \\
5 & 3 &  +  &   \ce{4f^13 5s^2 5p^5 ^3F}   & 24790 & 15003 & 35586 & 26215 & 22385 & 47695 & 21905 &  \\
6 & 0 &  +  &   \ce{4f^14 5s^2 5p^4 ^1S}   & 26393 & 63806 & 25088 & 26363 & 26090 & 26497 & 29810 &  \\
7 & 2 &  +  &   \ce{4f^13 5s^2 5p^5 ^1D}   & 26846 & 15362 & 33986 & 27655 & 23529 & 47483 & 23276 &  \\
8 & 2 &  +  &   \ce{4f^13 5s^2 5p^5 ^3D}   & 31473 & 21798 & 41766 & 32527 & 28953 & 54222 & 28112 &  \\
9 & 4 &  +  &   \ce{4f^13 5s^2 5p^5 ^3G}   & 38190 & 28550 & 48265 & 39073 & 35699 & 61013 & 34884 &  \\
10 & 1 &  +  &   \ce{4f^13 5s^2 5p^5 ^3D}   & 40416 & 30810 & 50131 & 40910 & 37781 & 63090 & 36497 &  \\
11 & 6 &  +  &   \ce{4f^12 5s^2 5p^6 ^3H}  & 56441 & 48067 & 77422 & 68158 & 50748 & 145685 & 56416 &  \\
12 & 4 &  +  &   \ce{4f^12 5s^2 5p^6 ^3F}  & 65552 & 56974 & 85853 & 76522 & 59469 & 154077 & 65008 &  \\
13 & 5 &  +  &   \ce{4f^12 5s^2 5p^6 ^3H}  & 72408 & 64602 & 93446 & 84302 & 67600 & 163611 & 73188 &  \\
14 & 4 &  +  &   \ce{4f^12 5s^2 5p^6 ^3H}  & 79883 & 72003 & 99172 & 91334 & 74974 & 171019 &     &  \\
15 & 3 &  +  &   \ce{4f^12 5s^2 5p^6 ^3F}  & 83359 & 75339 & 103309 & 94063 & 78231 & 174034 &     &  \\
16 & 2 &  +  &   \ce{4f^12 5s^2 5p^6 ^3F}  & 83682 & 75310 & 102875 & 92657 & 77822 & 172858 &     &  \\
17 & 1 &  +  &   \ce{4f^14 5s^2 5p^4 ^3P}  & 85630 & 123720 & 86025 & 86458 & 89174 & 86989 &     &  \\
18 & 4 &  +  &   \ce{4f^12 5s^2 5p^6 ^1G}  & 95159 & 87680 & 100606 & 106804 & 90941 & 187886 &     &  \\
19 & 2 &  +  &   \ce{4f^14 5s^2 5p^4 ^1D}  & 101529 & 144021 & 98607 & 100314 & 104751 & 102885 &     &  \\
20 & 2 &  +  &   \ce{4f^12 5s^2 5p^6 ^3P}  & 106797 & 98537 & 54576 & 115390 & 100800 & 196033 &     &  \\
21 & 3 &  +  &   \ce{4f^13 5s^2 5p^5 ^3D}  & 110930 & 100683 & 120460 & 111063 & 110651 & 135916 &     &  \\
22 & 4 &  +  &   \ce{4f^13 5s^2 5p^5 ^1G}  & 113533 & 103386 & 123174 & 113993 & 113422 & 138550 &     &  \\
23 & 6 &  +  &   \ce{4f^12 5s^2 5p^6 ^1I}  & 116027 & 107482 & 134862 & 125610 & 109762 & 204521 &     &  \\
24 & 0 &  +  &   \ce{4f^12 5s^2 5p^6 ^3P}  & 120387 & 112367 & 135796 & 127310 & 113437 & 207758 &     &  \\
25 & 1 &  +  &   \ce{4f^12 5s^2 5p^6 ^3P}  & 122862 & 111792 & 138962 & 130193 & 116590 & 211522 &     &  \\
26 & 3 &  +  &   \ce{4f^13 5s^2 5p^5 ^3G}  & 124043 & 113375 & 134019 & 124496 & 124629 & 149928 &     &  \\
27 & 2 &  +  &   \ce{4f^12 5s^2 5p^6 ^3P}  & 126047 & 115763 & 123973 & 133184 & 120467 & 216445 &     &  \\
28 & 2 &  +  &   \ce{4f^13 5s^2 5p^5 ^1D}  & 132931 & 119202 & 132495 & 138179 & 133867 & 160903 &     &  \\
29 & 0 &  +  &   \ce{4f^12 5s^2 5p^6 ^1S}  & 187736 & 175950 & 192803 & 181971 & 179945 & 276993 &     &  \\
30 & 0 &  +  &   \ce{4f^14 5s^2 5p^4 ^3P}  & 201002 & 237137 & 198587 & 208645 & 207545 & 200031 &     &  \\
31 & 2 &  -  &   \ce{4f^14 5s 5p^5 ^3P}  & 352893 & 352496 & 337528 & 330540 & 354339 & 360508 &     &  \\
32 & 1 &  -  &   \ce{4f^14 5s 5p^5 ^3P}  & 387715 & 387123 & 366821 & 359693 & 389383 & 395404 &     &  \\
33 & 4 &  -  &   \ce{4f^13 5s 5p^6 ^3F}  & 401019 & 387182 & 381935 & 372291 & 398449 & 453860 &     &  \\
34 & 3 &  -  &   \ce{4f^13 5s 5p^6 ^3F}  & 405276 & 391483 & 386293 & 376663 & 402921 & 458407 &     &  \\
35 & 2 &  -  &   \ce{4f^13 5s 5p^6 ^3F}  & 418363 & 404853 & 399400 & 389533 & 416830 & 472835 &     &  \\
36 & 3 &  -  &   \ce{4f^13 5s 5p^6 ^1F}  & 423720 & 410232 & 404822 & 395289 & 422426 & 478463 &     &  \\
37 & 0 &  -  &   \ce{4f^14 5s 5p^5 ^3P}  & 443061 & 440847 & 427980 & 420566 & 447765 & 452460 &     &  \\
38 & 1 &  -  &   \ce{4f^14 5s 5p^5 ^1P}  & 489350 & 487903 & 458918 & 451389 & 492939 & 498168 &     &  \\
\midrule
\multicolumn{12}{l}{Transitions (nm)}\\
8$\to$3   &      &  &      & 475.55 & 482.17 & 474.30 & 477.34 & 455.40 & 456.35 & 459.66 &  477.3$\pm$0.1\\
9$\to$4   &      &  &      & 426.92 & 434.05 & 432.96 & 434.79 & 404.59 & 404.62 & 420.84 &  431.8$\pm$0.1\\
13$\to$11   &    &  &      & 626.31 & 604.76 & 624.09 & 619.44 & 593.40 & 557.83 & 596.23 &  611.2$\pm$0.1\\
%
%
\end{longtable}

\section{Method of calculation}

Valence electron configurations associated with atomic levels studied in the present work are collected in Table~\ref{tab:conf}. We assumed that the core configuration is \ce{[Kr] 4d^10}. To specify the electronic configurations, we used the grouped forms, e.g., \ce{4f^{14\ldots12} 5s^2 5p^{4\ldots6}} means \ce{4f^14 5s^2 5p^4} and \ce{4f^13 5s^2 5p^5} and \ce{4f^12 5s^2 5p^6} together. 
Atomic levels are divided into four blocks corresponding to given configurations and level energies. 
The block 1 collects atomic levels close to the ground state; they are associated with \ce{4f^{14\ldots12} 5s^2 5p^{4\ldots6}} even configurations, with excitation energies below 30~eV. 
The block 2 is the first platform of odd states, associated with \ce{4f^{14\ldots13} 5s^1 5p^{5\ldots6}} configurations, with energies in the range of 40--60~eV. 
The block 3 is the second platform of odd states, with energies in the range of 40--90~eV. The block 3 configurations originate from the block 1 configurations by exciting an electron from the 4f or 5p subshell into the 5d subshell. 
The block 4 collects all other studied configurations.

\subsection{MCDHF method}

The calculations of the energy levels and radiative transition rates have been carried out by means of the \textsc{Grasp2018} \cite{FroeseFischer2018} code, based on the MCDHF method.
The methodology of MCDHF calculations performed in the present study is similar to that published earlier in many papers (see, e.g., \cite{Grant1984,Dyall1989,Grant2007}).
The effective Hamiltonian for an $N$-electron system is expressed by
\begin{equation}
H = \sum_{i=1}^{N} h_{D}(i) + \sum_{j>i=1}^{N} C_{ij},
\end{equation}
where $h_D(i)$ is the Dirac operator for the $i$th electron and the terms $C_{ij}$ account for the electron--electron interactions.
In general, the latter is a sum of the Coulomb interaction operator and the transverse Breit operator.
An atomic state function (ASF) with total angular momentum $J$ and parity $p$ is assumed in the form
\begin{equation}
\Psi_{s} (J^{p} ) = \sum_{m} c_{m} (s) \Phi ( \gamma_{m} J^{p} ),
\end{equation}
where $\Phi(\gamma_{m} J^{p})$ represents the configuration state functions (CSFs), $c_{m}(s)$ represents the configuration mixing coefficients for state $s$, and $\gamma_{m}$ represents all information required to define a certain CSF uniquely. 
The CSFs can be thought of as linear combinations of $N$-electron Slater determinants, which are antisymmetrised products of four-component Dirac orbital spinors:
\begin{equation}
\Phi(\gamma_m J^p) = 
\sum_i d_i 
\begin{vmatrix}
\psi_1(1) & \cdots & \psi_1(N)\\
\vdots & \ddots & \vdots \\
\psi_N(1) & \cdots & \psi_N(N)
\end{vmatrix}
\end{equation}
where $\psi_i$ is the one-electron wavefunction, defined as
\begin{equation}
\psi_{n, \kappa, j}=
\frac{1}{r}
\begin{pmatrix}
P_{n,\kappa}(r)\cdot \Omega_{\kappa, j}^{m_j}(\theta,\phi)\\[0.8ex]
i Q_{n,\kappa}(r)\cdot \Omega_{-\kappa, j}^{m_j}(\theta,\phi)
\end{pmatrix}
\end{equation}
where $\Omega_{\kappa, j}^{m_j}(\theta,\phi)$ is an angular two-component spinor and $P_{n,\kappa}(r)$ and $Q_{n,\kappa}(r)$ are large and small radial parts of the wavefunction, respectively.

In the \textsc{Grasp2018} code, the Breit interaction contribution to the energy is added perturbatively, after the radial part of the wavefunction has been optimised. 
In the present work the Breit term is calculated at the low-frequency limit. 
Also, two types of quantum electrodynamics (QED) corrections -- the self-energy (as the screened hydrogenic approximation \cite{McKenzie1980} of the data of Mohr and co-workers \cite{Mohr1992a}) and the vacuum polarisation (as the potential of Fullerton and Rinker \cite{Fullerton1976}) -- were included. 
The radiative transition rates for E1 and E2 transitions were calculated in both velocity (Coulomb) and length (Babushkin) \cite{Grant1974} gauges.

For comparative purposes we performed calculations by using the \fac code \cite{Gu2008}, based on the multiconfigurational Dirac--Hartree--Fock--Slater (MCDHFS) method. 
In general, the MCDHFS method is similar to the MCDHF method, in that they both refer to the effective Hamiltonian and multiconfigurational ASF. The main difference between the MCDHF and the MCDHFS methods is the approximation of the non-local Dirac--Hartree--Fock exchange potential by a local potential used in the latter. \fac uses an improved form of the local exchange potential (see \cite{Gu2008} for details).

\subsection{Probing the effect of electron correlation}

To determine the optimal computational scheme, one has to find the proper approach to calculate the optimal radial part of orbitals and to examine the effect of inter-electron correlation. 
Table~\ref{tab:Wix-conv} collects theoretical results obtained from various approaches for states assigned to electron configurations of blocks 1 and 2. 
We present results of our calculations performed by \grasp and \fac codes that are based on the relativistic multiconfigurational methods. However, the extensive CI calculations are performed only by using the \grasp code. 
Theoretical values obtained by Berengut et al. \cite{Berengut2011} are also listed in Table~\ref{tab:Wix-conv}. 

Both the Extended Average Level (EAL) and the Extended Optimal Level (EOL) schemes \cite{Grant1984} were used in order to calculate optimal orbitals. 
In the EOL scheme, the radial wavefunctions were optimised separately for each of the ASF subset related to given electron configurations in Table~\ref{tab:conf}. 
In the EAL scheme, the radial wavefunctions were optimised separately for each of the ASF subset related to all even configurations (block 1) or all odd electron configurations (block 2). 
The EOL and the EAL schemes were tested in the cases of both \grasp and \fac codes. 

The extensive CI calculations were preceded by determining the radial wavefunctions calculated within the self-consistent field process. 
The MCDHF-CI methodology used in this work is similar to the one presented in our previous papers \cite{Rzadkiewicz2018,Kozio2018}. 
The multireference (MR) set includes CSFs from configurations collected in Table~\ref{tab:conf}. 
For the EOL scheme, there are three MR sets for block 1 states -- \{\ce{4f^14 5s^2 5p^4}\}, \{\ce{4f^13 5s^2 5p^5}\}, and \{\ce{4f^12 5s^2 5p^6}\} -- and two MR sets for block 2 states -- \{\ce{4f^14 5s^1 5p^5}\}, and \{\ce{4f^13 5s^1 5p^6}\}. 
For the EAL scheme, there is one MR set for block 1 states -- \{\ce{4f^{14\ldots12} 5s^2 5p^{4\ldots6}}\} -- and one MR set for block 2 states -- \{\ce{4f^{14\ldots13} 5s^1 5p^{5\ldots6}}\}. 
We considered all possible single (S) and double (D) substitutions from the $4f$, $5s$, $5p$ occupied subshells into an active space (AS) of virtual orbitals. 
We used the ASs extended layer by layer.  
AS1 contains virtual orbitals with $n \le 5$ and $l \le 4$, the AS2 contains virtual orbitals with $n \le 6$ and $l \le 4$, and AS3 contains virtual orbitals with $n \le 7$ and $l \le 4$.  
In calculations, the highest layer had 1~733~659 CSFs for even states and 400~574 CSFs for odd states. 
When a new layer is added, the radial wavefunctions of the previous layers were frozen and only virtual orbitals introduced in a new layer were optimized. 

Our collisional-radiative simulations performed by the \fac code allow us to select a few transitions between low-lying states that can be observed in the visible spectral range. 
In order to verify the theoretical approaches, we compared the atomic transition calculations with available experimental results for \Wix obtained by Mita et al. \cite{Mita2017} in the visible range (400--620~nm). These experimental data are compared with MCDHF and FAC predictions at the bottom of Table~\ref{tab:Wix-conv}. 

\begin{figure}[!htb]
\centering
\includegraphics[width=0.8\linewidth]{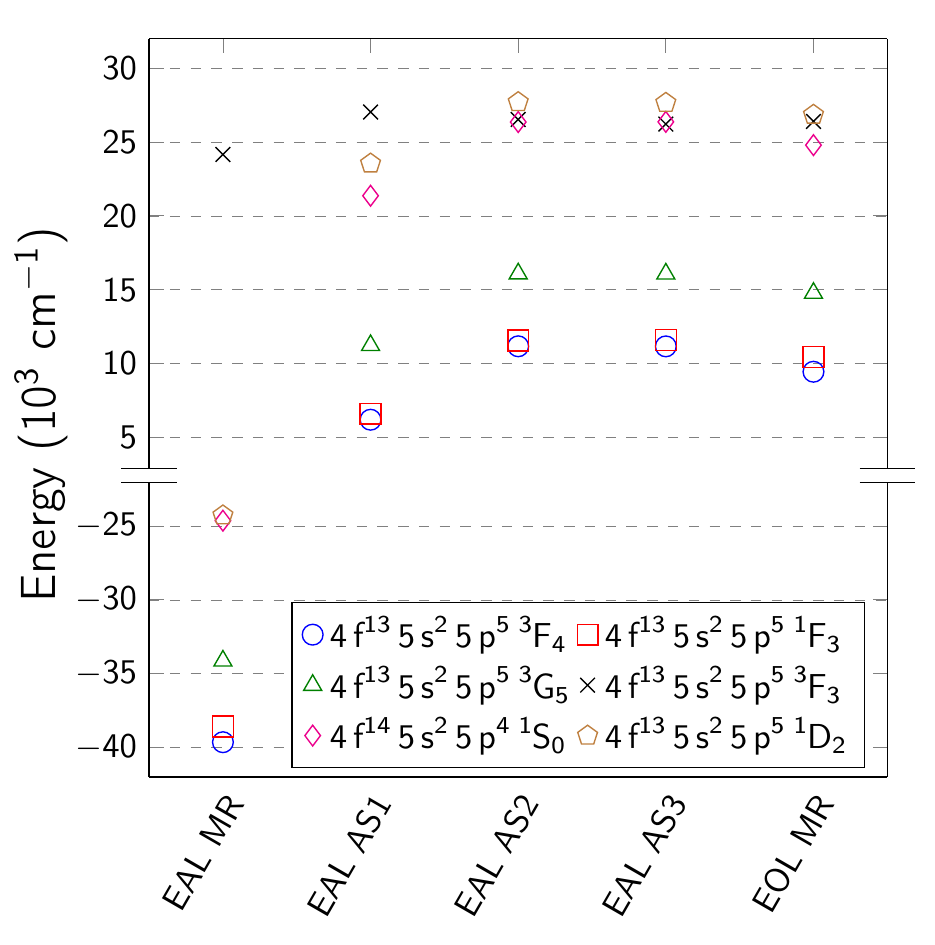}
\caption{Level energy convergence in CI calculation of first six excited states in \Wix.\label{fig:Wix-conv}}
\end{figure}

\begin{figure}[!htb]
\centering
\includegraphics[width=\linewidth]{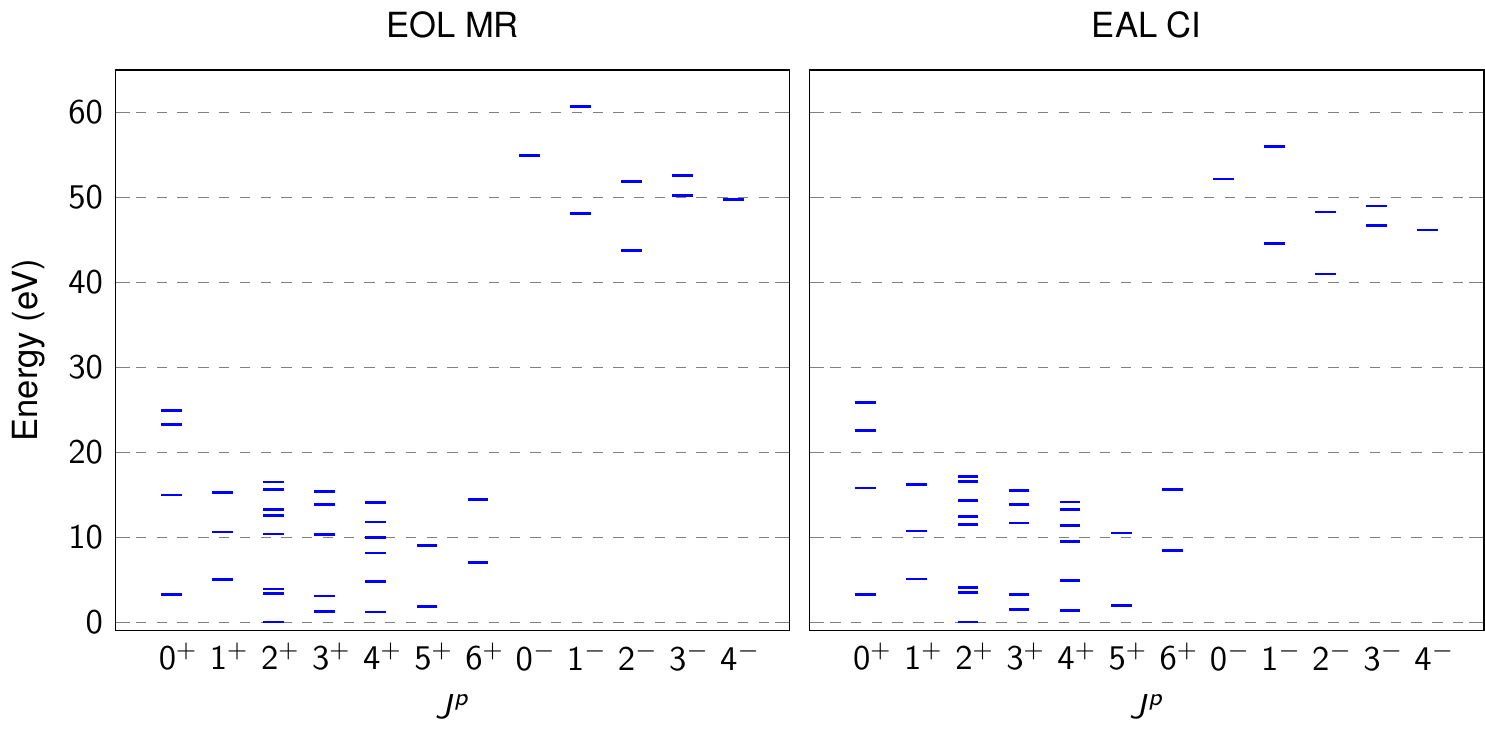}
\caption{Level energies calculated by EOL MR (left) and EAL CI AS3 (right) theoretical approaches for \Wix states assigned to blocks 1 and 2 electron configurations.\label{fig:Wix_levels_J_mrci}}
\end{figure}

Fig.~\ref{fig:Wix-conv} presents the convergence in CI calculation of the first six excited levels of \Wix. 
The level energies calculated by using only the MR set and the EOL scheme provide a fairly good approximation to the EAL CI energies at the AS3 stage. However, using the EOL scheme with CI calculations, one encounters problems related to mixing states with different radial wavefunctions. 
As shown in Fig.~\ref{fig:Wix-conv} and in Table~\ref{tab:Wix-conv}, the level energies calculated by using only the MR set and the EAL scheme are significantly lower than the final CI energies (except for the fourth excited state). It is because the groundstate is related to the \ce{4f^14 5s^2 5p^4} electron configuration (and the fourth excited state, too), while the first, second, third, fifth, and sixth excited states are related to the \ce{4f^13 5s^2 5p^5} electron configuration. Therefore, using the radial wavefunctions optimised in the EAL scheme for the mix of \ce{4f^14 5s^2 5p^4}, \ce{4f^13 5s^2 5p^5}, and \ce{4f^12 5s^2 5p^6} electron configurations may lead to incorrect results. The extensive CI procedure improves calculations of the level energies, and the convergence is reached at about the AS2--AS3 stage. 
It is worth noticing that both \fac-calculated and Berengut et al. \cite{Berengut2011} values reproduce experimental data less accurately than the \grasp calculations. 
A comparison between all level energies of states of block 1 and 2 calculated at EOL MR and EAL CI AS3 levels of theory is presented in Fig.~\ref{fig:Wix_levels_J_mrci}.

\begin{figure}[!htb]
\centering
\includegraphics[width=0.8\linewidth]{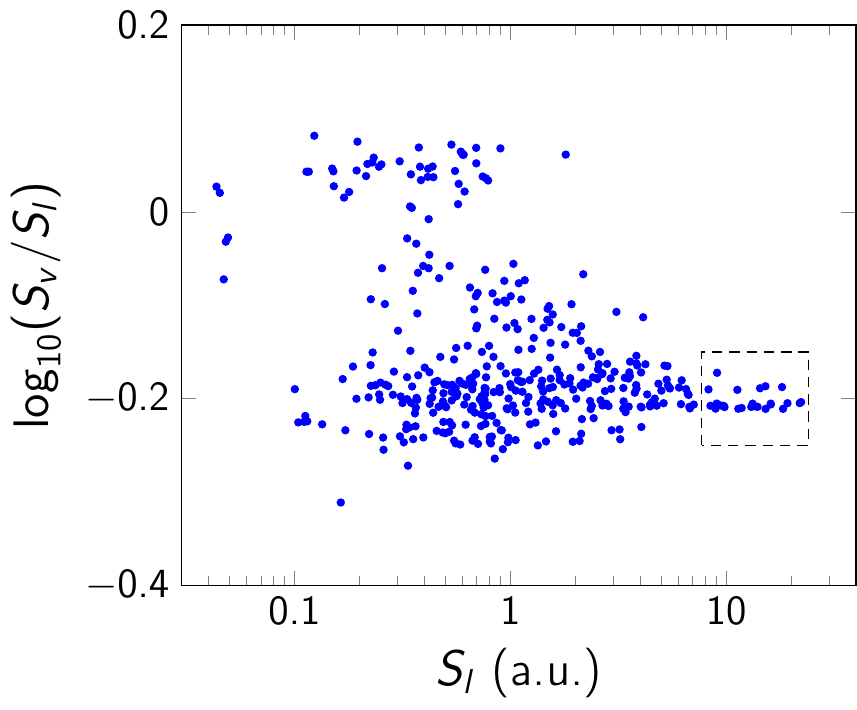}
\caption{Scatter plot of $\log_{10}(S_v/S_l)$ vs $S_l$ for selected high-intensity transitions between blocks 2 and 3 (upper levels) and block 1 (lower levels) in \Wix, calculated at the EOL MR level of theory. Dashed area indicates transitions selected for estimating transition rate uncertainty.\label{fig:Wix-tr-S-sel}}
\end{figure}

\begin{table*}[!htb]
\centering
\caption{High-intensity transitions selected for estimating transition rate uncertainty. Level numbering is according to Table~\ref{tab:Wix}.}\label{tab:tr-un}
\begin{tabular*}{\linewidth}{@{} ll @{\extracolsep{\fill}} ll @{\extracolsep{\fill}} crrc @{\extracolsep{\fill}} crrc @{\extracolsep{\fill}} c @{}}
\toprule
\multicolumn{2}{c}{Upper} & \multicolumn{2}{c}{Lower} & \multicolumn{4}{c}{\grasp} & \multicolumn{4}{c}{\fac} & Err (\%) \\
\cmidrule{1-2}\cmidrule{3-4}\cmidrule{5-8}\cmidrule{9-12}
No. & $J$ & No. & $J$ & $E$ (eV) & $S_l$ (a.u.) & $S_v$ (a.u.) & $\Delta_1$ (\%) & $E$ (eV) & $S_l$ (a.u.) & $S_v$ (a.u.) & $\Delta_2$ (\%) &  \\
\midrule
1320 & 7 & 23 & 6 & 66.8422 & 22.26 & 13.93 & 37.4 & 66.5834 & 21.61 & 13.76 & 2.2 & 37.5 \\
1093 & 7 & 11 & 6 & 66.9492 & 21.93 & 13.69 & 37.6 & 66.6997 & 20.77 & 13.20 & 4.6 & 37.9 \\
1191 & 6 & 13 & 5 & 66.9382 & 19.24 & 12.01 & 37.6 & 66.6766 & 18.70 & 11.88 & 2.1 & 37.7 \\
1335 & 6 & 23 & 6 & 67.9086 & 18.97 & 11.54 & 39.2 & 67.6295 & 18.39 & 11.39 & 2.4 & 39.3 \\
1124 & 6 & 11 & 6 & 67.5305 & 18.35 & 11.30 & 38.5 & 67.2884 & 17.37 & 10.88 & 4.7 & 38.7 \\
722 & 6 & 4 & 5 & 64.5451 & 18.16 & 11.79 & 35.1 & 64.3964 & 17.17 & 11.83 & 3.1 & 35.2 \\
1287 & 5 & 18 & 4 & 67.0001 & 16.11 & 10.03 & 37.8 & 66.7299 & 15.63 & 9.90 & 2.3 & 37.8 \\
1322 & 5 & 23 & 6 & 67.0037 & 16.09 & 10.03 & 37.7 & 66.7715 & 15.58 & 9.88 & 2.5 & 37.8 \\
1222 & 5 & 13 & 5 & 67.5388 & 15.24 & 9.38 & 38.5 & 67.2414 & 7.94 & 4.98 & 47.5 & 61.1 \\
849 & 5 & 9 & 4 & 64.4799 & 15.21 & 9.90 & 34.9 & 64.3292 & 14.37 & 9.92 & 3.3 & 35.1 \\
1271 & 5 & 22 & 4 & 63.9441 & 14.36 & 9.30 & 35.3 & 63.7944 & 13.51 & 9.27 & 3.7 & 35.4 \\
1241 & 5 & 14 & 4 & 67.1383 & 13.98 & 8.64 & 38.2 & 66.9337 & 12.73 & 8.00 & 8.3 & 39.1 \\
1168 & 5 & 12 & 4 & 67.2101 & 13.57 & 8.40 & 38.1 & 67.0087 & 11.33 & 7.10 & 16.1 & 41.3 \\
1092 & 5 & 11 & 6 & 66.9138 & 13.27 & 8.27 & 37.7 & 66.6939 & 13.97 & 8.87 & -6.1 & 38.2 \\
1292 & 4 & 18 & 4 & 67.3077 & 13.07 & 8.08 & 38.2 & 67.0500 & 12.64 & 7.96 & 2.6 & 38.2 \\
1249 & 4 & 14 & 4 & 67.4585 & 11.80 & 7.27 & 38.3 & 67.2477 & 10.39 & 6.52 & 11.3 & 40.0 \\
1262 & 4 & 15 & 3 & 67.3805 & 11.39 & 7.01 & 38.4 & 67.1446 & 10.58 & 6.62 & 6.6 & 39.0 \\
1297 & 4 & 26 & 3 & 64.1623 & 11.27 & 7.27 & 35.5 & 64.0192 & 10.64 & 7.27 & 3.4 & 35.7 \\
1311 & 3 & 20 & 2 & 67.3340 & 9.83 & 6.08 & 38.2 & 67.0792 & 9.50 & 5.98 & 2.7 & 38.3 \\
1293 & 3 & 18 & 4 & 67.3088 & 9.74 & 6.04 & 38.0 & 67.0926 & 9.30 & 5.87 & 3.9 & 38.2 \\
1166 & 3 & 12 & 4 & 67.1545 & 9.58 & 5.94 & 38.0 & 66.9350 & 9.64 & 6.11 & -1.5 & 38.0 \\
1341 & 3 & 27 & 2 & 67.2268 & 9.24 & 5.74 & 37.8 & 66.9685 & 9.17 & 5.80 & 0.0 & 37.8 \\
1105 & 3 & 19 & 2 & 61.5451 & 9.08 & 6.11 & 32.7 & 61.6587 & 8.38 & 5.66 & 7.6 & 33.6 \\
1200 & 4 & 13 & 5 & 67.0365 & 9.04 & 5.64 & 37.7 & 66.8425 & 10.91 & 6.91 & -21.4 & 43.3 \\
1266 & 3 & 16 & 2 & 67.4339 & 8.94 & 5.50 & 38.4 & 67.2006 & 8.31 & 5.21 & 6.4 & 39.0 \\
1260 & 3 & 15 & 3 & 67.3123 & 8.46 & 5.25 & 38.0 & 67.1070 & 7.45 & 4.70 & 11.3 & 39.6 \\
720 & 5 & 2 & 4 & 65.0786 & 8.29 & 5.35 & 35.4 & 64.9720 & 6.76 & 4.34 & 18.6 & 40.0 \\
\multicolumn{4}{@{}l}{Average} &  &  &  & 37.3$\pm$1.5 &  &  &  & 5.5$\pm$11.0 & 39.0$\pm$4.9 \\
\bottomrule
\end{tabular*}
\end{table*}

For levels from blocks 3 and 4, we used the EOL MR approach, because the large number of these levels precludes performing extensive CI calculations with reasonable quality in reasonable time. 
Note that the sizes of the MR sets for blocks 3 and 4 are much larger than the ones for blocks 1 and 2. Thus, we assume that a significant part of the electron correlation (static correlation) for blocks 3 and 4 states is taken into account at the EOL MR level of theory. 
For the sake of consistency, the final data set of level energies is presented for numbers obtained for the EOL MR approach. 
The electric dipole (E1), magnetic dipole (M1), electric quadrupole (E2), and magnetic quadrupole (M2) radiative transitions between all states from blocks 1, 2, 3, and 4 were calculated at the EOL MR level of theory. 
The total number of $\approx 3.17\times10^8$ transitions was considered. 
The obtained transition rates have been used to calculate the radiative lifetimes of given states. 

\subsection{Estimation of transition rate uncertainties}

The theoretical uncertainties of transition rates were estimated according to the procedure described by Kramida \cite{Kramida2013}. 
Fig.~\ref{fig:Wix-tr-S-sel} presents the scatter plot of $\log_{10}(S_v/S_l)$ vs $S_l$ (where $S_l$ and $S_v$ are line strengths calculated in length and velocity gauges, respectively) for selected high-intensity transitions between blocks 2 and 3 (upper levels) and block 1 (lower levels) in \Wix, calculated at the EOL MR level of theory by using \grasp code. 
Among these transitions, the subset of high-intensity transitions (with $S_l \geq 8~\text{a.u.}$) has been selected for estimating transition rate uncertainty. These transitions are listed in Table~\ref{tab:tr-un}, together with complementary calculations of these transitions performed by using \fac code. 
The factor $\Delta_1=(S_l-S_v)/S_l$ evaluates the difference in rates calculated in length and velocity gauges. 
The factor $\Delta_2=[(S_l+S_v)^\text{Gr}-(S_l+S_v)^\text{Fac}]/(S_l+S_v)^\text{Gr}$ evaluates the difference between \grasp and \fac results. 
The factor $\text{Err}=\sqrt{\Delta_1^2+\Delta_2^2}$ provides estimation for theoretical uncertainty of transition rates. The obtained value of uncertainty has been extended to all transitions presented in this work.

\begin{figure*}[!h]
\centering
\includegraphics[width=\linewidth]{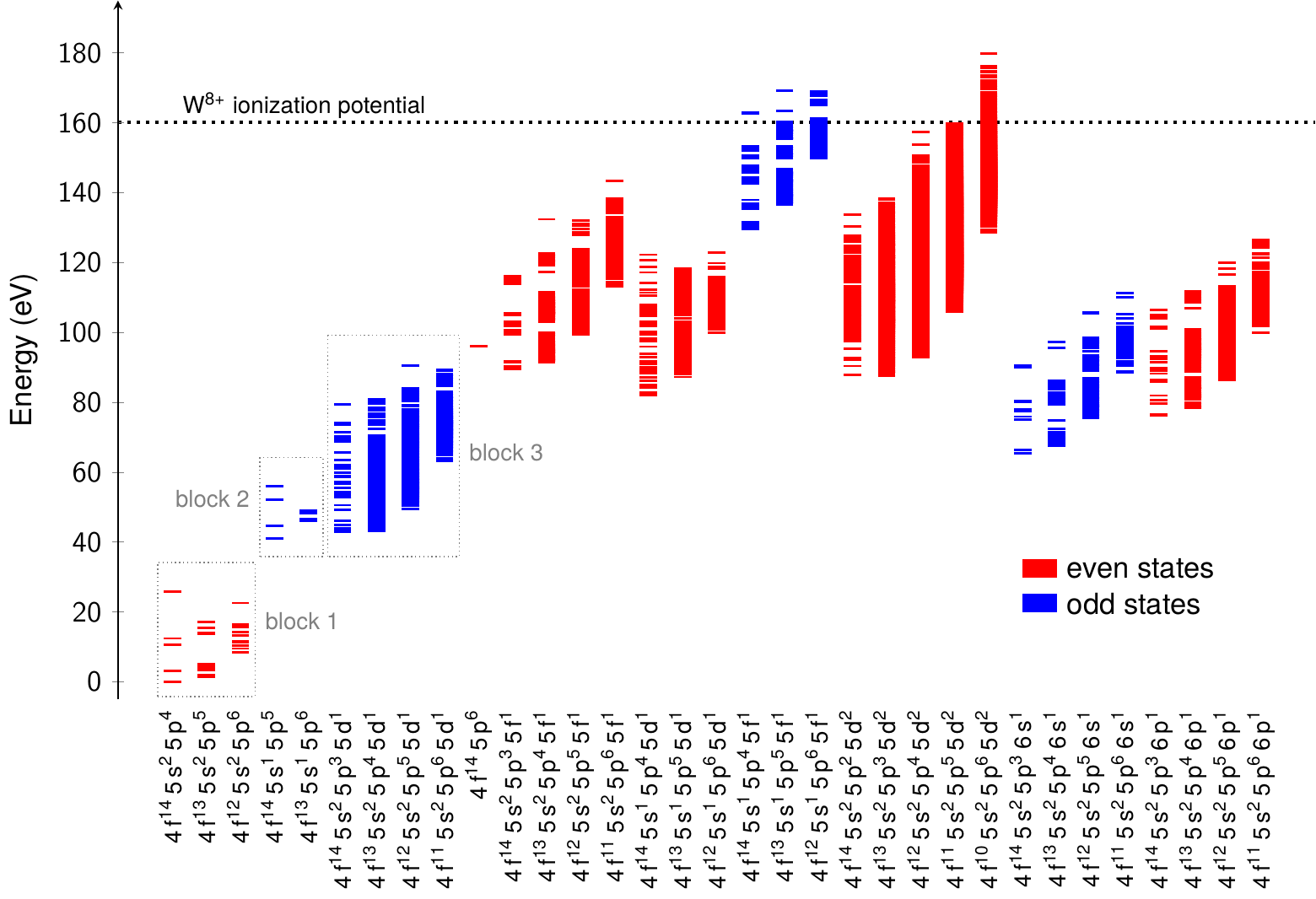}
\caption{Energy level diagram of \Wix.\label{Wix-lev}}
\end{figure*}

\section{Results}

\subsection{Ionisation potential of \Wix}

The ionisation potential (IP) for \Wix, calculated at the EAL AS2 level of theory as a difference between the ground states for \Wix and \Wx ions, is 160.11~eV. 
Note that IP calculated at the EOL MR level of theory is 158.53~eV, which allows us to estimate the effect of electron correlation on IP to be about 1.6~eV. 
The present calculated IP value is very close to the value 160.2$\pm$1.2~eV obtained by Kramida and Reader \cite{Kramida2006} and to the number 156.125~eV from Beiersdorfer et al. \cite{Beiersdorfer2012}. 
The IP obtained from the binding energies presented by Rodrigues et al. \cite{Rodrigues2004} is 176~eV. This value is incorrect, because Rodrigues et al. \cite{Rodrigues2004} did not assign the proper configurations to the ground levels of \Wix and \Wx.

\subsection{Level structure of \Wix}

Table \ref{tab:Wix} presents atomic levels assigned to block 1, 2, and 3 configurations of \Wix. 
The energy of atomic levels assigned to block 4 configurations of {\Wix} are collected in the Supplementary Material table. 
The theoretical uncertainties of energy levels have been roughly estimated, by comparing various theoretical approaches, to be $\approx$5000~\cmm for \Wix levels below 50~000~\cmm and $\approx$10~000~\cmm for levels above it. 
The energy level diagram of all studied configurations of \Wix is presented in Fig.~\ref{Wix-lev}. 

For the \Wix ion, Kramida and Shirai \cite{Kramida2009} suggested that the groundstate is \ce{4f^14 5s^2 5p^4 ^3P_2} and the first excited level (probably \ce{4f^13 5s^2 5p^5 ^3F_4}) is located about 14~000~\cmm above the ground state. 
Berengut et al. \cite{Berengut2011} calculated 16 levels related to \ce{4f^14 5s^2 5p^4}, \ce{4f^13 5s^2 5p^5}, and \ce{4f^12 5s^2 5p^6} electron configurations. They confirmed that the groundstate is \ce{4f^14 5s^2 5p^4 ^3P_2} and reported that the first excited level \ce{4f^13 5s^2 5p^5 ^3F_4} is located 6075~\cmm above the groundstate, while the second excited level \ce{4f^13 5s^2 5p^5 ^3G_3} is located 6357~\cmm above the groundstate. 
The present calculations seem to confirm that the ground state is \ce{4f^14 5s^2 5p^4 ^3P_2} and the first excited state has a dominant contribution of \ce{4f^13 5s^2 5p^5 ^3F_4}. The relative energy of the first excited state is predicted to be 9440~\cmm above the ground state. This value is between the numbers provided by Kramida and Shirai \cite{Kramida2009} and by Berengut et al. \cite{Berengut2011}. From the present calculations, the second excited state is dominated by the \ce{4f^13 5s^2 5p^5 ^1F_3} contribution in its composition, and it is located 10~445~\cmm above the ground state. So, the \ce{^3F_4} and the \ce{^1F_3} levels are competing for being the first excited state.

\subsection{Selected transitions of \Wix}

Selected high-intensity E1 transitions, with $A_l \geq 10^{10}~\text{s}^{-1}$, between block 1 and block 2 and 3 configurations of \Wix are presented in Table~\ref{tab:Wix_tr}. 
The theoretical uncertainties of calculated transition wavelengths are roughly estimated to be $\approx$2~\AA{}. 
The theoretical uncertainties of transition rates are estimated to be $\approx$40\%. 
Because many transitions are close to each other, it is hardly possible to identify our theoretical transition wavelengths with the ones measured by Ryabtsev et al. \cite{Ryabtsev2015}. More theoretical work in the future is needed for clarification of this issue.

Selected high-intensity M1 and E2 transitions, with $A_M \geq 1~\text{s}^{-1}$ or $A_l \geq 1~\text{s}^{-1}$, within block 1 levels of \Wix are presented in Table~\ref{tab:Wix_trf}. All M1 and E2 transitions within block 1 levels of \Wix are collected in the Supplementary Material table. Energies and rates of the E1, M1, E2, and M2 transitions of decaying metastable ($\tau_l > 10^{-6}$~s) states of blocks 2 and 3 are also collected in the Supplementary Material table. The theoretical uncertainties of these E1, M1, E2, and M2 transition rates are estimated to be $\approx$40\%.

\section*{Supplementary Material}
\noindent
Six supplementary files can be seen in the zip which contains \texttt{supplement-table1.txt} (computer readable form of Table~1), \texttt{supplement-table2.txt} (computer readable form of Table~2), \texttt{supplement-table3.txt} (computer readable form of Table~3), \texttt{supplement-levels-blocks1234.txt} (energies, compositions, and radiative lifetimes of all levels: blocks 1, 2, 3, and 4), \texttt{supplement-transitions-metastable-block1.txt} (energies and rates of M1 and E2 transitions of decaying metastable states of block 1), \texttt{supplement-transitions-metastable-blocks23.txt} (energies and rates of E1, M1, E2, and M2 transitions of decaying metastable states of blocks 2 and 3).

\ack

This work was partly supported by the Polish Ministry of Science and Higher Education within the framework of the scientific financial resources in the years 2016--2020 allocated for the realisation of the international co-financed project. This work has been carried out within the framework of the EUROfusion Consortium and has received funding from the Euratom Research and Training Programme 2014--2020 under Grant Agreement No. 633053. The views and opinions expressed herein do not necessarily reflect those of the European Commission.

The authors wish to thank Nobuyuki Nakamura for providing the unpublished experimental wavelengths.


%

\clearpage

\TableExplanation

\section*{Table 1.}
\begin{tabular}{@{}p{0.5in}p{6in}@{}}
$J$		& Total momentum of the state\\
$p$		& Parity of the state\\
$E$		& Energy of atomic level relative to the groundstate (\cmm/eV)\\
$\tau_l$   & Radiative lifetime of the state calculated in length (Babushkin) gauge (s)\\
$\tau_v$   & Radiative lifetime of the state calculated in velocity (Coulomb) gauge (s)\\
\end{tabular}

\medskip

\noindent
For each level the state composition in the terms of $LS$-coupling scheme is provided. Up to three $LS$ terms with contribution larger than 10\% are listed.

\medskip

\section*{Tables 2 and 3.}
\begin{tabular}{@{}p{0.5in}p{6in}@{}}
$J$		& Total momentum of the state\\
$\lambda$		& Transition wavelength (\AA)\\
$E$		& Transition energy (eV)\\
$A_l$   & Transition rate for electric-type transitions, calculated in length (Babushkin) gauge (s$^{-1}$)\\
$A_v$   & Transition rate for electric-type transitions, calculated in velocity (Coulomb) gauge (s$^{-1}$)\\
$A_M$   & Transition rate for magnetic-type transitions (s$^{-1}$)\\
\end{tabular}

\medskip

\noindent
Upper and lower state numbers are related to the states in Table~\ref{tab:Wix}.

\datatables 

\setlength{\LTcapwidth}{\linewidth}
\setlength{\LTleft}{0pt}
\setlength{\LTright}{0pt} 
\setlength{\tabcolsep}{0.5\tabcolsep}
\renewcommand{\arraystretch}{1.0}

\begin{longtable}{@{\extracolsep{\fill}} rcc >{\raggedright\arraybackslash}p{0.6\linewidth} rrrr}
\caption{Atomic levels of \Wix.\label{tab:Wix}}
No. & $J$ & $p$ & $LS$-composition & $E$ (\cmm) & $E$ (eV) & $\tau_l$ (s) & $\tau_v$ (s) \\
\midrule
\endfirsthead
\multicolumn{8}{l}{Table \ref{tab:Wix} (continued)}\\
\midrule
No. & $J$ & $p$ & $LS$-composition & $E$ (\cmm) & $E$ (eV) & $\tau_l$ (s) & $\tau_v$ (s) \\
\midrule
\endhead
\midrule
\endfoot
\bottomrule
\endlastfoot
1 & 2 &  +    &  0.76~$4f^{14}(^{1}_{1}S)\,5s^{2}\,5p^{4}(^{3}_{2}P)~^{3}P$ + 0.24~$4f^{14}(^{1}_{1}S)\,5s^{2}\,5p^{4}(^{1}_{2}D)~^{1}D$  & 0 & 0.0000 & -- & -- \\
2 & 4 &  +    &  0.81~$4f^{13}(^{2}_{1}F)\,5s^{2}\,5p^{5}~^{3}F$ + 0.10~$4f^{13}(^{2}_{1}F)\,5s^{2}\,5p^{5}~^{1}G$ + 0.10~$4f^{13}(^{2}_{1}F)\,5s^{2}\,5p^{5}~^{3}G$  & 9440 & 1.1704 & 1.37E+03 & 6.22E+01 \\
3 & 3 &  +    &  0.59~$4f^{13}(^{2}_{1}F)\,5s^{2}\,5p^{5}~^{1}F$ + 0.28~$4f^{13}(^{2}_{1}F)\,5s^{2}\,5p^{5}~^{3}D$ + 0.09~$4f^{13}(^{2}_{1}F)\,5s^{2}\,5p^{5}~^{3}F$  & 10445 & 1.2950 & 5.47E+02 & 6.67E+01 \\
4 & 5 &  +    &  1.00~$4f^{13}(^{2}_{1}F)\,5s^{2}\,5p^{5}~^{3}G$                                             & 14767 & 1.8309 & 3.15E+00 & 3.15E+00 \\
5 & 3 &  +    &  0.62~$4f^{13}(^{2}_{1}F)\,5s^{2}\,5p^{5}~^{3}F$ + 0.17~$4f^{13}(^{2}_{1}F)\,5s^{2}\,5p^{5}~^{3}G$ + 0.15~$4f^{13}(^{2}_{1}F)\,5s^{2}\,5p^{5}~^{1}F$  & 24791 & 3.0737 & 1.33E-02 & 1.33E-02 \\
6 & 0 &  +    &  0.52~$4f^{14}(^{1}_{1}S)\,5s^{2}\,5p^{4}(^{1}_{0}S)~^{1}S$ + 0.48~$4f^{14}(^{1}_{1}S)\,5s^{2}\,5p^{4}(^{3}_{2}P)~^{3}P$  & 26393 & 3.2723 & 4.18E-01 & 4.67E+00 \\
7 & 2 &  +    &  0.57~$4f^{13}(^{2}_{1}F)\,5s^{2}\,5p^{5}~^{1}D$ + 0.24~$4f^{13}(^{2}_{1}F)\,5s^{2}\,5p^{5}~^{3}F$ + 0.19~$4f^{13}(^{2}_{1}F)\,5s^{2}\,5p^{5}~^{3}D$  & 26847 & 3.3286 & 2.01E+00 & 2.17E+00 \\
8 & 2 &  +    &  0.56~$4f^{13}(^{2}_{1}F)\,5s^{2}\,5p^{5}~^{3}D$ + 0.44~$4f^{13}(^{2}_{1}F)\,5s^{2}\,5p^{5}~^{3}F$  & 31473 & 3.9022 & 9.85E-03 & 9.83E-03 \\
\end{longtable}

\clearpage

\setlength{\LTcapwidth}{\linewidth}
\setlength{\LTleft}{0pt}
\setlength{\LTright}{0pt} 
\setlength{\tabcolsep}{0.5\tabcolsep}
\renewcommand{\arraystretch}{1.0}

\begin{longtable}{@{\extracolsep{\fill}} llll  rrrr}
\caption{High-intensity E1 transitions of \Wix.\label{tab:Wix_tr}}\\
\multicolumn{2}{c}{Upper level} & \multicolumn{2}{c}{Lower level} &&&&\\
\cmidrule{1-2}\cmidrule{3-4}
No. & $J$ & No. & $J$ & $\lambda$ (\AA) & $E$ (eV) & $A_l$ (s$^{-1}$) & $A_v$ (s$^{-1}$) \\
\midrule
\endfirsthead
\multicolumn{8}{l}{Table \ref{tab:Wix_tr} (continued)}\\
\midrule
\multicolumn{2}{c}{Upper level} & \multicolumn{2}{c}{Lower level} &&&&\\
\cmidrule{1-2}\cmidrule{3-4}
No. & $J$ & No. & $J$ & $\lambda$ (\AA) & $E$ (eV) & $A_l$ (s$^{-1}$) & $A_v$ (s$^{-1}$) \\
\midrule
\endhead
\midrule
\endfoot
\bottomrule
\endlastfoot
1354 & 5 & 23 & 6 & 170.793 & 72.5934 & 1.46E+10 & 1.28E+10 \\
1349 & 0 & 25 & 1 & 176.658 & 70.1833 & 1.77E+10 & 1.64E+10 \\
1021 & 3 & 5 & 3 & 178.531 & 69.4468 & 1.86E+10 & 1.15E+10 \\
1339 & 1 & 20 & 2 & 178.563 & 69.4343 & 2.04E+10 & 1.19E+10 \\
920 & 4 & 2 & 4 & 178.605 & 69.4183 & 1.43E+10 & 8.70E+09 \\
918 & 3 & 3 & 3 & 179.110 & 69.2225 & 3.35E+10 & 2.06E+10 \\
1334 & 1 & 20 & 2 & 179.651 & 69.0138 & 1.21E+10 & 7.23E+09 \\
1295 & 1 & 17 & 1 & 180.063 & 68.8559 & 2.86E+10 & 1.82E+10 \\
920 & 4 & 4 & 5 & 180.320 & 68.7579 & 2.18E+10 & 1.40E+10 \\
1262 & 4 & 13 & 5 & 180.371 & 68.7383 & 2.01E+10 & 1.20E+10 \\
876 & 4 & 2 & 4 & 180.572 & 68.6620 & 1.21E+10 & 7.56E+09 \\
1342 & 5 & 23 & 6 & 180.580 & 68.6588 & 1.31E+10 & 1.29E+10 \\
1344 & 1 & 24 & 0 & 181.144 & 68.4449 & 1.30E+10 & 7.78E+09 \\
1011 & 2 & 8 & 2 & 181.259 & 68.4017 & 2.39E+10 & 1.55E+10 \\
1345 & 3 & 27 & 2 & 181.317 & 68.3798 & 4.44E+10 & 2.59E+10 \\
1168 & 5 & 11 & 6 & 181.423 & 68.3397 & 2.37E+10 & 1.43E+10 \\
1344 & 1 & 25 & 1 & 181.960 & 68.1380 & 9.24E+10 & 5.59E+10 \\
1213 & 5 & 12 & 4 & 181.972 & 68.1335 & 1.39E+10 & 8.13E+09 \\
876 & 4 & 4 & 5 & 182.326 & 68.0015 & 2.33E+10 & 1.48E+10 \\
1335 & 6 & 23 & 6 & 182.575 & 67.9086 & 4.86E+11 & 2.95E+11 \\
1266 & 3 & 14 & 4 & 182.585 & 67.9049 & 3.49E+10 & 2.12E+10 \\
1021 & 3 & 9 & 4 & 182.907 & 67.7855 & 5.39E+10 & 3.46E+10 \\
1339 & 1 & 24 & 0 & 183.004 & 67.7494 & 1.24E+10 & 7.49E+09 \\
1260 & 3 & 14 & 4 & 183.021 & 67.7432 & 4.59E+10 & 2.82E+10 \\
1344 & 1 & 27 & 2 & 183.021 & 67.7431 & 3.77E+11 & 2.30E+11 \\
1343 & 2 & 27 & 2 & 183.291 & 67.6434 & 2.79E+10 & 1.74E+10 \\
1128 & 5 & 11 & 6 & 183.412 & 67.5986 & 1.04E+10 & 1.06E+10 \\
1222 & 5 & 13 & 5 & 183.574 & 67.5388 & 4.54E+11 & 2.79E+11 \\
1124 & 6 & 11 & 6 & 183.597 & 67.5305 & 4.62E+11 & 2.84E+11 \\
1267 & 1 & 16 & 2 & 183.705 & 67.4911 & 1.96E+11 & 1.21E+11 \\
1249 & 4 & 14 & 4 & 183.793 & 67.4585 & 4.28E+11 & 2.64E+11 \\
1340 & 2 & 25 & 1 & 183.819 & 67.4492 & 1.54E+11 & 9.47E+10 \\
1339 & 1 & 25 & 1 & 183.837 & 67.4425 & 3.63E+11 & 2.23E+11 \\
1266 & 3 & 16 & 2 & 183.860 & 67.4339 & 4.16E+11 & 2.56E+11 \\
1262 & 4 & 15 & 3 & 184.006 & 67.3805 & 4.12E+11 & 2.53E+11 \\
1173 & 4 & 12 & 4 & 184.024 & 67.3740 & 2.45E+11 & 1.51E+11 \\
1331 & 5 & 23 & 6 & 184.049 & 67.3647 & 1.02E+10 & 1.03E+10 \\
1311 & 3 & 20 & 2 & 184.133 & 67.3340 & 4.56E+11 & 2.82E+11 \\
1334 & 1 & 24 & 0 & 184.147 & 67.3289 & 4.35E+11 & 2.69E+11 \\
1310 & 1 & 20 & 2 & 184.153 & 67.3267 & 4.39E+11 & 2.71E+11 \\
1322 & 5 & 22 & 4 & 184.190 & 67.3131 & 1.04E+10 & 8.58E+09 \\
1260 & 3 & 15 & 3 & 184.193 & 67.3123 & 3.92E+11 & 2.43E+11 \\
1293 & 3 & 18 & 4 & 184.202 & 67.3088 & 4.51E+11 & 2.80E+11 \\
1292 & 4 & 18 & 4 & 184.205 & 67.3077 & 4.71E+11 & 2.91E+11 \\
1011 & 2 & 10 & 1 & 184.246 & 67.2928 & 3.88E+10 & 2.54E+10 \\
1213 & 5 & 13 & 5 & 184.271 & 67.2835 & 1.37E+10 & 8.47E+09 \\
1257 & 2 & 15 & 3 & 184.278 & 67.2812 & 4.37E+11 & 2.70E+11 \\
1261 & 1 & 16 & 2 & 184.300 & 67.2731 & 2.58E+11 & 1.60E+11 \\
1337 & 2 & 25 & 1 & 184.357 & 67.2523 & 3.09E+11 & 1.91E+11 \\
1366 & 1 & 29 & 0 & 184.390 & 67.2404 & 4.79E+11 & 2.97E+11 \\
904 & 3 & 5 & 3 & 184.418 & 67.2300 & 2.53E+10 & 1.64E+10 \\
1341 & 3 & 27 & 2 & 184.427 & 67.2268 & 4.26E+11 & 2.65E+11 \\
804 & 4 & 2 & 4 & 184.447 & 67.2193 & 1.99E+10 & 1.28E+10 \\
1168 & 5 & 12 & 4 & 184.473 & 67.2101 & 3.98E+11 & 2.47E+11 \\
918 & 3 & 7 & 2 & 184.531 & 67.1889 & 3.54E+10 & 2.29E+10 \\
1243 & 3 & 14 & 4 & 184.577 & 67.1720 & 1.31E+11 & 8.14E+10 \\
801 & 3 & 2 & 4 & 184.580 & 67.1710 & 2.87E+10 & 1.87E+10 \\
1166 & 3 & 12 & 4 & 184.625 & 67.1545 & 4.41E+11 & 2.73E+11 \\
1208 & 4 & 13 & 5 & 184.649 & 67.1459 & 9.56E+10 & 5.93E+10 \\
1106 & 5 & 11 & 6 & 184.665 & 67.1402 & 3.46E+10 & 2.16E+10 \\
1241 & 5 & 14 & 4 & 184.670 & 67.1383 & 4.09E+11 & 2.53E+11 \\
1253 & 2 & 16 & 2 & 184.697 & 67.1285 & 3.97E+11 & 2.47E+11 \\
896 & 2 & 5 & 3 & 184.734 & 67.1151 & 4.33E+10 & 2.83E+10 \\
804 & 4 & 3 & 3 & 184.790 & 67.0947 & 1.19E+10 & 7.90E+09 \\
1100 & 5 & 11 & 6 & 184.809 & 67.0877 & 2.16E+10 & 1.35E+10 \\
1307 & 2 & 20 & 2 & 184.815 & 67.0857 & 4.54E+11 & 2.83E+11 \\
1163 & 4 & 12 & 4 & 184.844 & 67.0751 & 2.58E+10 & 1.62E+10 \\
1340 & 2 & 27 & 2 & 184.901 & 67.0543 & 2.84E+11 & 1.77E+11 \\
1339 & 1 & 27 & 2 & 184.920 & 67.0476 & 8.43E+10 & 5.23E+10 \\
801 & 3 & 3 & 3 & 184.923 & 67.0464 & 3.51E+10 & 2.08E+10 \\
1200 & 4 & 13 & 5 & 184.951 & 67.0365 & 3.22E+11 & 2.01E+11 \\
1236 & 3 & 14 & 4 & 184.997 & 67.0197 & 2.35E+11 & 1.46E+11 \\
1159 & 4 & 12 & 4 & 185.006 & 67.0163 & 1.61E+11 & 1.01E+11 \\
1322 & 5 & 23 & 6 & 185.041 & 67.0037 & 4.68E+11 & 2.92E+11 \\
1287 & 5 & 18 & 4 & 185.051 & 67.0001 & 4.68E+11 & 2.91E+11 \\
1093 & 7 & 11 & 6 & 185.191 & 66.9492 & 4.66E+11 & 2.91E+11 \\
1191 & 6 & 13 & 5 & 185.222 & 66.9382 & 4.72E+11 & 2.94E+11 \\
1092 & 5 & 11 & 6 & 185.289 & 66.9138 & 3.84E+11 & 2.39E+11 \\
1295 & 1 & 19 & 2 & 185.370 & 66.8846 & 1.18E+10 & 7.01E+09 \\
1337 & 2 & 27 & 2 & 185.445 & 66.8574 & 1.49E+11 & 9.34E+10 \\
1320 & 7 & 23 & 6 & 185.488 & 66.8422 & 4.71E+11 & 2.95E+11 \\
1332 & 0 & 25 & 1 & 185.623 & 66.7934 & 4.73E+11 & 2.96E+11 \\
1242 & 2 & 15 & 3 & 185.858 & 66.7092 & 2.10E+10 & 1.32E+10 \\
1243 & 3 & 16 & 2 & 185.881 & 66.7010 & 1.57E+10 & 9.78E+09 \\
1242 & 2 & 16 & 2 & 185.969 & 66.6691 & 4.69E+10 & 2.92E+10 \\
772 & 4 & 2 & 4 & 185.992 & 66.6611 & 1.71E+10 & 1.07E+10 \\
1224 & 4 & 14 & 4 & 186.046 & 66.6416 & 2.14E+10 & 1.33E+10 \\
1334 & 1 & 27 & 2 & 186.087 & 66.6271 & 2.03E+10 & 1.28E+10 \\
1236 & 3 & 15 & 3 & 186.194 & 66.5887 & 6.19E+10 & 3.86E+10 \\
804 & 4 & 4 & 5 & 186.277 & 66.5589 & 1.01E+10 & 6.79E+09 \\
772 & 4 & 3 & 3 & 186.340 & 66.5365 & 1.78E+10 & 1.16E+10 \\
1234 & 2 & 16 & 2 & 186.347 & 66.5341 & 1.12E+10 & 1.18E+10 \\
1173 & 4 & 13 & 5 & 186.375 & 66.5240 & 1.51E+10 & 9.56E+09 \\
1214 & 3 & 14 & 4 & 186.674 & 66.4174 & 1.38E+10 & 8.77E+09 \\
904 & 3 & 8 & 2 & 186.719 & 66.4015 & 4.82E+10 & 3.18E+10 \\
1061 & 6 & 11 & 6 & 186.719 & 66.4015 & 1.02E+10 & 6.47E+09 \\
901 & 1 & 8 & 2 & 186.740 & 66.3939 & 2.39E+10 & 1.69E+10 \\
1213 & 5 & 14 & 4 & 186.845 & 66.3566 & 5.98E+10 & 3.90E+10 \\
896 & 2 & 8 & 2 & 187.043 & 66.2866 & 4.08E+10 & 2.50E+10 \\
1224 & 4 & 15 & 3 & 187.257 & 66.2107 & 1.99E+10 & 1.32E+10 \\
1106 & 5 & 12 & 4 & 187.825 & 66.0106 & 1.50E+10 & 9.77E+09 \\
770 & 5 & 4 & 5 & 187.959 & 65.9634 & 1.73E+11 & 1.14E+11 \\
1214 & 3 & 16 & 2 & 188.007 & 65.9464 & 1.63E+10 & 1.09E+10 \\
920 & 4 & 9 & 4 & 188.272 & 65.8538 & 8.52E+10 & 5.66E+10 \\
831 & 2 & 5 & 3 & 188.272 & 65.8537 & 4.05E+10 & 2.62E+10 \\
1208 & 4 & 15 & 3 & 188.460 & 65.7881 & 2.33E+10 & 1.57E+10 \\
1092 & 5 & 12 & 4 & 188.471 & 65.7842 & 2.31E+10 & 1.48E+10 \\
1091 & 4 & 12 & 4 & 188.483 & 65.7802 & 1.80E+10 & 1.15E+10 \\
1026 & 7 & 11 & 6 & 188.638 & 65.7262 & 1.59E+10 & 1.72E+10 \\
1310 & 1 & 24 & 0 & 188.880 & 65.6418 & 1.01E+10 & 6.51E+09 \\
704 & 2 & 1 & 2 & 188.951 & 65.6171 & 6.34E+10 & 4.27E+10 \\
831 & 2 & 7 & 2 & 189.004 & 65.5988 & 1.00E+10 & 6.65E+09 \\
1315 & 0 & 25 & 1 & 189.050 & 65.5829 & 1.31E+10 & 1.39E+10 \\
904 & 3 & 9 & 4 & 189.091 & 65.5687 & 3.33E+10 & 2.28E+10 \\
735 & 3 & 2 & 4 & 189.137 & 65.5524 & 4.31E+10 & 2.80E+10 \\
737 & 4 & 3 & 3 & 189.138 & 65.5523 & 3.53E+10 & 2.27E+10 \\
1020 & 5 & 11 & 6 & 189.258 & 65.5108 & 1.05E+10 & 1.13E+10 \\
809 & 3 & 5 & 3 & 189.361 & 65.4750 & 1.12E+11 & 7.02E+10 \\
1314 & 1 & 25 & 1 & 189.361 & 65.4750 & 1.52E+10 & 1.61E+10 \\
735 & 3 & 3 & 3 & 189.498 & 65.4278 & 1.43E+11 & 8.94E+10 \\
1310 & 1 & 25 & 1 & 189.767 & 65.3349 & 2.34E+10 & 1.53E+10 \\
805 & 2 & 5 & 3 & 189.777 & 65.3316 & 1.34E+10 & 9.16E+09 \\
804 & 4 & 5 & 3 & 189.822 & 65.3161 & 7.11E+10 & 4.61E+10 \\
901 & 1 & 10 & 1 & 189.912 & 65.2851 & 2.26E+11 & 1.48E+11 \\
801 & 3 & 5 & 3 & 189.962 & 65.2677 & 1.57E+10 & 9.86E+09 \\
842 & 1 & 8 & 2 & 190.176 & 65.1946 & 1.66E+11 & 1.10E+11 \\
896 & 2 & 10 & 1 & 190.225 & 65.1777 & 9.97E+10 & 6.67E+10 \\
1309 & 3 & 26 & 3 & 190.279 & 65.1591 & 1.96E+11 & 1.24E+11 \\
876 & 4 & 9 & 4 & 190.460 & 65.0974 & 1.73E+11 & 1.14E+11 \\
724 & 2 & 3 & 3 & 190.477 & 65.0913 & 1.71E+11 & 1.13E+11 \\
720 & 5 & 2 & 4 & 190.515 & 65.0786 & 2.21E+11 & 1.43E+11 \\
805 & 2 & 7 & 2 & 190.520 & 65.0767 & 2.85E+11 & 1.86E+11 \\
831 & 2 & 8 & 2 & 190.671 & 65.0252 & 1.61E+11 & 1.03E+11 \\
801 & 3 & 7 & 2 & 190.707 & 65.0128 & 1.12E+11 & 7.50E+10 \\
1166 & 3 & 16 & 2 & 191.019 & 64.9067 & 1.04E+10 & 6.82E+09 \\
1283 & 4 & 21 & 3 & 191.199 & 64.8455 & 1.24E+11 & 8.02E+10 \\
1225 & 5 & 18 & 4 & 191.283 & 64.8172 & 1.09E+10 & 1.19E+10 \\
711 & 4 & 2 & 4 & 191.345 & 64.7961 & 1.61E+11 & 1.04E+11 \\
1303 & 2 & 26 & 3 & 191.457 & 64.7582 & 5.77E+10 & 3.77E+10 \\
772 & 4 & 5 & 3 & 191.458 & 64.7578 & 1.94E+11 & 1.26E+11 \\
1361 & 1 & 29 & 0 & 191.489 & 64.7473 & 3.33E+10 & 3.65E+10 \\
711 & 4 & 3 & 3 & 191.714 & 64.6715 & 1.75E+11 & 1.13E+11 \\
809 & 3 & 8 & 2 & 191.788 & 64.6465 & 2.21E+11 & 1.44E+11 \\
1279 & 3 & 21 & 3 & 191.835 & 64.6307 & 2.75E+11 & 1.75E+11 \\
1150 & 1 & 16 & 2 & 191.923 & 64.6010 & 1.45E+10 & 1.60E+10 \\
1305 & 1 & 27 & 2 & 191.924 & 64.6006 & 1.09E+10 & 1.20E+10 \\
776 & 1 & 7 & 2 & 192.052 & 64.5577 & 3.19E+11 & 2.06E+11 \\
722 & 6 & 4 & 5 & 192.089 & 64.5451 & 3.99E+11 & 2.59E+11 \\
1264 & 1 & 20 & 2 & 192.115 & 64.5364 & 1.85E+10 & 2.05E+10 \\
1283 & 4 & 22 & 4 & 192.156 & 64.5228 & 2.11E+11 & 1.35E+11 \\
707 & 3 & 2 & 4 & 192.158 & 64.5221 & 1.88E+10 & 1.24E+10 \\
864 & 0 & 10 & 1 & 192.184 & 64.5131 & 3.97E+11 & 2.58E+11 \\
1140 & 1 & 16 & 2 & 192.231 & 64.4975 & 1.42E+10 & 1.58E+10 \\
849 & 5 & 9 & 4 & 192.283 & 64.4799 & 3.94E+11 & 2.56E+11 \\
801 & 3 & 8 & 2 & 192.405 & 64.4392 & 1.49E+10 & 9.45E+09 \\
720 & 5 & 4 & 5 & 192.468 & 64.4182 & 1.11E+11 & 7.10E+10 \\
1318 & 1 & 28 & 2 & 192.551 & 64.3905 & 3.57E+11 & 2.29E+11 \\
1279 & 3 & 22 & 4 & 192.797 & 64.3080 & 7.59E+10 & 4.98E+10 \\
1203 & 5 & 18 & 4 & 192.900 & 64.2738 & 1.07E+10 & 1.19E+10 \\
1201 & 3 & 18 & 4 & 193.026 & 64.2320 & 1.02E+10 & 1.14E+10 \\
1297 & 4 & 26 & 3 & 193.235 & 64.1623 & 3.52E+11 & 2.27E+11 \\
711 & 4 & 4 & 5 & 193.316 & 64.1357 & 2.17E+10 & 1.45E+10 \\
750 & 3 & 5 & 3 & 193.373 & 64.1164 & 6.16E+10 & 4.09E+10 \\
842 & 1 & 10 & 1 & 193.466 & 64.0857 & 1.32E+11 & 8.51E+10 \\
1309 & 3 & 28 & 2 & 193.552 & 64.0571 & 1.53E+11 & 9.98E+10 \\
747 & 2 & 5 & 3 & 193.586 & 64.0459 & 1.35E+11 & 8.98E+10 \\
943 & 6 & 11 & 6 & 193.625 & 64.0333 & 1.49E+10 & 1.68E+10 \\
688 & 5 & 2 & 4 & 193.771 & 63.9848 & 5.53E+10 & 3.63E+10 \\
776 & 1 & 8 & 2 & 193.774 & 63.9840 & 6.16E+10 & 4.01E+10 \\
1298 & 1 & 27 & 2 & 193.879 & 63.9491 & 2.13E+10 & 2.40E+10 \\
1271 & 5 & 22 & 4 & 193.895 & 63.9441 & 3.63E+11 & 2.35E+11 \\
996 & 3 & 12 & 4 & 193.908 & 63.9397 & 1.22E+10 & 1.38E+10 \\
831 & 2 & 10 & 1 & 193.979 & 63.9163 & 8.75E+10 & 5.68E+10 \\
1258 & 2 & 21 & 3 & 194.139 & 63.8637 & 3.61E+11 & 2.33E+11 \\
685 & 3 & 2 & 4 & 194.158 & 63.8574 & 1.34E+11 & 8.91E+10 \\
809 & 3 & 9 & 4 & 194.291 & 63.8137 & 1.58E+10 & 1.08E+10 \\
737 & 4 & 5 & 3 & 194.413 & 63.7736 & 1.08E+11 & 7.16E+10 \\
685 & 3 & 3 & 3 & 194.538 & 63.7328 & 1.75E+10 & 1.15E+10 \\
680 & 4 & 2 & 4 & 194.655 & 63.6943 & 5.02E+10 & 3.41E+10 \\
1303 & 2 & 28 & 2 & 194.772 & 63.6562 & 3.01E+11 & 1.95E+11 \\
804 & 4 & 9 & 4 & 194.776 & 63.6548 & 2.99E+10 & 1.89E+10 \\
740 & 1 & 7 & 2 & 194.855 & 63.6288 & 7.05E+10 & 4.69E+10 \\
680 & 4 & 3 & 3 & 195.037 & 63.5697 & 9.23E+10 & 6.22E+10 \\
645 & 3 & 1 & 2 & 195.078 & 63.5563 & 1.00E+11 & 6.88E+10 \\
674 & 5 & 2 & 4 & 195.306 & 63.4821 & 9.98E+10 & 6.72E+10 \\
735 & 3 & 7 & 2 & 195.576 & 63.3943 & 9.78E+10 & 6.47E+10 \\
805 & 2 & 10 & 1 & 195.576 & 63.3943 & 1.22E+10 & 7.98E+09 \\
1263 & 6 & 23 & 6 & 195.591 & 63.3897 & 3.76E+10 & 4.34E+10 \\
677 & 2 & 3 & 3 & 195.637 & 63.3748 & 1.44E+11 & 9.66E+10 \\
750 & 3 & 8 & 2 & 195.905 & 63.2879 & 4.96E+10 & 3.33E+10 \\
747 & 2 & 8 & 2 & 196.124 & 63.2174 & 4.84E+10 & 3.31E+10 \\
772 & 4 & 9 & 4 & 196.499 & 63.0965 & 1.67E+10 & 1.06E+10 \\
724 & 2 & 7 & 2 & 196.620 & 63.0578 & 4.77E+10 & 3.06E+10 \\
740 & 1 & 8 & 2 & 196.628 & 63.0552 & 9.67E+10 & 6.52E+10 \\
680 & 4 & 4 & 5 & 196.694 & 63.0339 & 8.31E+10 & 5.71E+10 \\
1074 & 2 & 17 & 1 & 196.830 & 62.9905 & 1.13E+11 & 7.68E+10 \\
664 & 3 & 2 & 4 & 196.916 & 62.9632 & 1.80E+10 & 1.26E+10 \\
664 & 3 & 3 & 3 & 197.306 & 62.8386 & 2.39E+10 & 1.72E+10 \\
674 & 5 & 4 & 5 & 197.359 & 62.8217 & 6.11E+10 & 4.14E+10 \\
1246 & 7 & 23 & 6 & 197.503 & 62.7758 & 1.22E+10 & 1.43E+10 \\
768 & 0 & 10 & 1 & 197.833 & 62.6710 & 1.29E+10 & 1.21E+10 \\
750 & 3 & 9 & 4 & 198.517 & 62.4551 & 8.83E+10 & 6.18E+10 \\
701 & 1 & 6 & 0 & 199.142 & 62.2591 & 1.97E+11 & 1.40E+11 \\
737 & 4 & 9 & 4 & 199.613 & 62.1123 & 3.82E+10 & 2.59E+10 \\
747 & 2 & 10 & 1 & 199.625 & 62.1085 & 4.26E+10 & 2.98E+10 \\
685 & 3 & 5 & 3 & 200.122 & 61.9542 & 2.88E+10 & 2.07E+10 \\
740 & 1 & 10 & 1 & 200.148 & 61.9463 & 1.57E+10 & 1.07E+10 \\
585 & 2 & 1 & 2 & 200.221 & 61.9237 & 1.52E+10 & 1.14E+10 \\
1240 & 1 & 25 & 1 & 200.650 & 61.7914 & 1.03E+10 & 1.25E+10 \\
685 & 3 & 7 & 2 & 200.949 & 61.6993 & 1.96E+10 & 1.36E+10 \\
574 & 1 & 1 & 2 & 201.055 & 61.6668 & 8.99E+10 & 6.73E+10 \\
1105 & 3 & 19 & 2 & 201.453 & 61.5451 & 3.21E+11 & 2.16E+11 \\
608 & 3 & 2 & 4 & 202.100 & 61.3479 & 1.21E+10 & 8.58E+09 \\
559 & 3 & 1 & 2 & 202.198 & 61.3181 & 1.81E+11 & 1.24E+11 \\
664 & 3 & 5 & 3 & 203.053 & 61.0600 & 5.44E+10 & 4.22E+10 \\
1074 & 2 & 19 & 2 & 203.189 & 61.0192 & 1.26E+11 & 8.91E+10 \\
543 & 3 & 1 & 2 & 203.390 & 60.9588 & 3.75E+10 & 2.67E+10 \\
593 & 3 & 2 & 4 & 203.768 & 60.8456 & 3.21E+10 & 2.58E+10 \\
657 & 2 & 5 & 3 & 203.874 & 60.8142 & 4.56E+10 & 3.65E+10 \\
593 & 3 & 3 & 3 & 204.187 & 60.7210 & 5.07E+10 & 3.99E+10 \\
529 & 1 & 1 & 2 & 204.353 & 60.6717 & 1.30E+10 & 6.34E+09 \\
588 & 3 & 3 & 3 & 204.391 & 60.6603 & 1.26E+10 & 9.79E+09 \\
587 & 4 & 3 & 3 & 204.450 & 60.6427 & 1.48E+10 & 1.06E+10 \\
579 & 4 & 2 & 4 & 204.489 & 60.6313 & 2.53E+10 & 1.90E+10 \\
567 & 4 & 2 & 4 & 205.525 & 60.3254 & 1.80E+10 & 1.46E+10 \\
1112 & 3 & 22 & 4 & 206.069 & 60.1664 & 1.90E+10 & 1.93E+10 \\
587 & 4 & 4 & 5 & 206.273 & 60.1069 & 2.23E+10 & 1.78E+10 \\
657 & 2 & 8 & 2 & 206.689 & 59.9857 & 4.61E+10 & 3.75E+10 \\
579 & 4 & 4 & 5 & 206.741 & 59.9709 & 2.97E+10 & 2.51E+10 \\
499 & 1 & 1 & 2 & 206.755 & 59.9668 & 4.00E+10 & 3.50E+10 \\
493 & 2 & 1 & 2 & 207.032 & 59.8866 & 1.16E+10 & 1.01E+10 \\
567 & 4 & 4 & 5 & 207.800 & 59.6650 & 2.35E+10 & 1.99E+10 \\
900 & 1 & 17 & 1 & 207.861 & 59.6477 & 1.53E+11 & 1.14E+11 \\
1205 & 1 & 28 & 2 & 207.960 & 59.6191 & 1.28E+10 & 1.32E+10 \\
668 & 0 & 10 & 1 & 208.646 & 59.4233 & 1.01E+10 & 1.06E+10 \\
664 & 3 & 9 & 4 & 208.732 & 59.3987 & 3.29E+10 & 2.90E+10 \\
444 & 2 & 1 & 2 & 209.998 & 59.0408 & 8.41E+10 & 6.70E+10 \\
864 & 0 & 17 & 1 & 210.473 & 58.9074 & 1.02E+10 & 8.66E+09 \\
657 & 2 & 10 & 1 & 210.582 & 58.8769 & 1.83E+10 & 1.65E+10 \\
593 & 3 & 7 & 2 & 211.262 & 58.6875 & 1.13E+10 & 1.04E+10 \\
578 & 3 & 7 & 2 & 212.096 & 58.4567 & 1.34E+10 & 1.46E+10 \\
1001 & 4 & 21 & 3 & 212.258 & 58.4119 & 1.75E+10 & 1.91E+10 \\
1142 & 3 & 28 & 2 & 212.280 & 58.4060 & 1.75E+10 & 1.87E+10 \\
835 & 0 & 17 & 1 & 212.490 & 58.3484 & 1.37E+11 & 1.14E+11 \\
568 & 1 & 7 & 2 & 212.921 & 58.2300 & 1.50E+10 & 1.64E+10 \\
1066 & 4 & 26 & 3 & 213.480 & 58.0777 & 1.79E+10 & 1.94E+10 \\
1107 & 1 & 28 & 2 & 215.015 & 57.6629 & 1.79E+10 & 1.17E+10 \\
531 & 2 & 5 & 3 & 215.230 & 57.6055 & 1.00E+10 & 1.12E+10 \\
603 & 2 & 10 & 1 & 216.375 & 57.3005 & 1.75E+10 & 1.95E+10 \\
351 & 2 & 1 & 2 & 216.869 & 57.1701 & 3.29E+10 & 2.69E+10 \\
1021 & 3 & 26 & 3 & 216.979 & 57.1411 & 1.02E+11 & 6.78E+10 \\
340 & 1 & 1 & 2 & 217.532 & 56.9958 & 2.75E+10 & 2.39E+10 \\
1011 & 2 & 26 & 3 & 217.805 & 56.9244 & 1.52E+11 & 1.04E+11 \\
920 & 4 & 21 & 3 & 218.147 & 56.8352 & 8.32E+10 & 5.73E+10 \\
530 & 1 & 8 & 2 & 218.380 & 56.7745 & 1.51E+10 & 1.73E+10 \\
918 & 3 & 21 & 3 & 218.421 & 56.7640 & 5.27E+10 & 3.49E+10 \\
499 & 1 & 6 & 0 & 218.688 & 56.6945 & 2.41E+10 & 2.08E+10 \\
989 & 3 & 26 & 3 & 219.300 & 56.5364 & 1.67E+10 & 1.92E+10 \\
920 & 4 & 22 & 4 & 219.393 & 56.5124 & 7.59E+10 & 5.11E+10 \\
918 & 3 & 22 & 4 & 219.670 & 56.4412 & 1.46E+11 & 9.99E+10 \\
892 & 3 & 21 & 3 & 219.985 & 56.3602 & 1.60E+10 & 1.86E+10 \\
394 & 5 & 4 & 5 & 220.638 & 56.1935 & 1.54E+10 & 1.81E+10 \\
458 & 3 & 5 & 3 & 220.949 & 56.1143 & 1.01E+10 & 1.19E+10 \\
876 & 4 & 21 & 3 & 221.089 & 56.0788 & 3.19E+10 & 2.23E+10 \\
1021 & 3 & 28 & 2 & 221.246 & 56.0391 & 1.13E+11 & 7.76E+10 \\
1318 & 1 & 30 & 0 & 221.595 & 55.9507 & 2.07E+10 & 1.94E+10 \\
1011 & 2 & 28 & 2 & 222.105 & 55.8224 & 5.61E+10 & 3.63E+10 \\
876 & 4 & 22 & 4 & 222.369 & 55.7561 & 2.88E+10 & 1.90E+10 \\
530 & 1 & 10 & 1 & 222.731 & 55.6656 & 1.20E+10 & 1.42E+10 \\
704 & 2 & 17 & 1 & 225.424 & 55.0004 & 1.27E+11 & 8.78E+10 \\
904 & 3 & 26 & 3 & 225.737 & 54.9242 & 3.30E+10 & 1.96E+10 \\
701 & 1 & 17 & 1 & 225.776 & 54.9147 & 1.94E+10 & 1.15E+10 \\
896 & 2 & 26 & 3 & 226.210 & 54.8093 & 3.98E+10 & 2.62E+10 \\
841 & 6 & 23 & 6 & 226.644 & 54.7044 & 1.80E+10 & 1.01E+10 \\
804 & 4 & 21 & 3 & 226.927 & 54.6362 & 3.14E+10 & 1.98E+10 \\
801 & 3 & 21 & 3 & 227.128 & 54.5879 & 1.99E+10 & 1.14E+10 \\
1295 & 1 & 30 & 0 & 227.279 & 54.5515 & 2.21E+11 & 1.55E+11 \\
876 & 4 & 26 & 3 & 227.690 & 54.4530 & 1.65E+10 & 9.80E+09 \\
1261 & 1 & 29 & 0 & 228.029 & 54.3721 & 1.47E+10 & 8.20E+09 \\
804 & 4 & 22 & 4 & 228.275 & 54.3135 & 1.27E+10 & 7.20E+09 \\
818 & 6 & 23 & 6 & 228.369 & 54.2912 & 1.18E+10 & 6.91E+09 \\
801 & 3 & 22 & 4 & 228.478 & 54.2651 & 1.84E+10 & 1.18E+10 \\
854 & 1 & 25 & 1 & 229.156 & 54.1047 & 1.24E+10 & 7.84E+09 \\
772 & 4 & 21 & 3 & 229.269 & 54.0780 & 1.37E+10 & 8.72E+09 \\
870 & 3 & 27 & 2 & 229.338 & 54.0619 & 1.05E+10 & 6.43E+09 \\
802 & 6 & 23 & 6 & 229.711 & 53.9740 & 1.03E+10 & 5.87E+09 \\
904 & 3 & 28 & 2 & 230.358 & 53.8222 & 2.50E+10 & 1.52E+10 \\
901 & 1 & 28 & 2 & 230.391 & 53.8147 & 1.82E+10 & 1.07E+10 \\
792 & 6 & 23 & 6 & 230.401 & 53.8124 & 1.04E+10 & 5.86E+09 \\
618 & 5 & 13 & 5 & 230.514 & 53.7860 & 1.39E+10 & 7.74E+09 \\
665 & 1 & 16 & 2 & 230.532 & 53.7817 & 3.23E+10 & 1.82E+10 \\
772 & 4 & 22 & 4 & 230.646 & 53.7552 & 1.26E+10 & 7.66E+09 \\
1239 & 1 & 29 & 0 & 230.774 & 53.7253 & 1.87E+10 & 1.10E+10 \\
770 & 5 & 22 & 4 & 230.806 & 53.7179 & 3.65E+10 & 2.19E+10 \\
896 & 2 & 28 & 2 & 230.851 & 53.7073 & 1.11E+10 & 5.92E+09 \\
532 & 6 & 11 & 6 & 230.914 & 53.6929 & 4.09E+10 & 2.33E+10 \\
659 & 2 & 15 & 3 & 231.069 & 53.6568 & 2.33E+10 & 1.31E+10 \\
697 & 5 & 18 & 4 & 231.180 & 53.6310 & 1.26E+10 & 6.88E+09 \\
656 & 3 & 15 & 3 & 231.547 & 53.5461 & 1.28E+10 & 7.29E+09 \\
693 & 4 & 18 & 4 & 231.554 & 53.5444 & 3.80E+10 & 2.16E+10 \\
656 & 3 & 16 & 2 & 231.720 & 53.5060 & 1.16E+10 & 7.18E+09 \\
570 & 4 & 12 & 4 & 231.865 & 53.4726 & 1.77E+10 & 1.01E+10 \\
840 & 3 & 27 & 2 & 231.914 & 53.4614 & 1.30E+10 & 7.32E+09 \\
817 & 1 & 25 & 1 & 232.025 & 53.4356 & 2.14E+10 & 1.23E+10 \\
654 & 2 & 16 & 2 & 232.120 & 53.4138 & 2.22E+10 & 1.27E+10 \\
731 & 3 & 20 & 2 & 232.151 & 53.4069 & 3.39E+10 & 1.92E+10 \\
811 & 2 & 25 & 1 & 232.226 & 53.3895 & 1.18E+10 & 6.94E+09 \\
837 & 1 & 27 & 2 & 232.338 & 53.3638 & 1.39E+10 & 7.96E+09 \\
650 & 4 & 15 & 3 & 232.742 & 53.2711 & 2.16E+10 & 1.32E+10 \\
630 & 5 & 14 & 4 & 232.907 & 53.2333 & 3.11E+10 & 1.80E+10 \\
726 & 2 & 20 & 2 & 233.021 & 53.2072 & 1.72E+10 & 1.02E+10 \\
598 & 6 & 13 & 5 & 233.144 & 53.1792 & 3.95E+10 & 2.31E+10 \\
824 & 1 & 27 & 2 & 233.245 & 53.1563 & 1.64E+10 & 9.44E+09 \\
557 & 3 & 12 & 4 & 233.402 & 53.1205 & 1.11E+10 & 6.43E+09 \\
506 & 7 & 11 & 6 & 233.420 & 53.1163 & 3.13E+10 & 1.83E+10 \\
798 & 2 & 25 & 1 & 233.640 & 53.0663 & 1.13E+10 & 6.44E+09 \\
704 & 2 & 19 & 2 & 233.804 & 53.0290 & 5.47E+10 & 4.11E+10 \\
550 & 5 & 12 & 4 & 233.867 & 53.0147 & 2.28E+10 & 1.38E+10 \\
804 & 4 & 26 & 3 & 233.886 & 53.0104 & 1.34E+10 & 8.41E+09 \\
501 & 5 & 11 & 6 & 233.932 & 53.0000 & 2.80E+10 & 1.59E+10 \\
500 & 7 & 11 & 6 & 233.975 & 52.9905 & 1.30E+10 & 7.71E+09 \\
735 & 3 & 21 & 3 & 234.068 & 52.9693 & 1.47E+10 & 9.71E+09 \\
718 & 1 & 20 & 2 & 234.103 & 52.9614 & 2.64E+10 & 1.53E+10 \\
701 & 1 & 19 & 2 & 234.183 & 52.9433 & 1.19E+10 & 9.58E+09 \\
676 & 3 & 18 & 4 & 234.517 & 52.8680 & 3.66E+10 & 2.13E+10 \\
582 & 4 & 13 & 5 & 234.520 & 52.8673 & 1.43E+10 & 8.25E+09 \\
752 & 5 & 23 & 6 & 234.530 & 52.8650 & 1.51E+10 & 8.61E+09 \\
751 & 7 & 23 & 6 & 234.533 & 52.8644 & 4.24E+10 & 2.50E+10 \\
675 & 5 & 18 & 4 & 234.544 & 52.8619 & 3.06E+10 & 1.84E+10 \\
616 & 3 & 14 & 4 & 234.573 & 52.8552 & 2.19E+10 & 1.24E+10 \\
544 & 3 & 12 & 4 & 234.645 & 52.8390 & 1.10E+10 & 6.54E+09 \\
724 & 2 & 21 & 3 & 235.565 & 52.6328 & 2.33E+10 & 1.44E+10 \\
842 & 1 & 28 & 2 & 235.642 & 52.6154 & 1.48E+10 & 9.41E+09 \\
760 & 1 & 24 & 0 & 236.148 & 52.5028 & 1.14E+10 & 6.58E+09 \\
153 & 3 & 1 & 2 & 236.237 & 52.4831 & 1.35E+10 & 1.42E+10 \\
565 & 4 & 13 & 5 & 236.332 & 52.4618 & 1.25E+10 & 7.37E+09 \\
768 & 0 & 25 & 1 & 236.390 & 52.4489 & 2.06E+10 & 1.22E+10 \\
720 & 5 & 22 & 4 & 237.642 & 52.1727 & 1.04E+10 & 6.77E+09 \\
750 & 3 & 26 & 3 & 239.303 & 51.8107 & 1.56E+10 & 1.11E+10 \\
237 & 3 & 5 & 3 & 239.926 & 51.6761 & 1.30E+10 & 7.71E+09 \\
737 & 4 & 26 & 3 & 240.896 & 51.4679 & 1.81E+10 & 1.19E+10 \\
595 & 0 & 17 & 1 & 240.938 & 51.4590 & 1.69E+10 & 1.87E+10 \\
658 & 1 & 19 & 2 & 241.560 & 51.3264 & 2.66E+10 & 2.94E+10 \\
172 & 5 & 4 & 5 & 241.603 & 51.3174 & 2.24E+10 & 1.40E+10 \\
153 & 3 & 2 & 4 & 241.625 & 51.3126 & 1.07E+10 & 6.22E+09 \\
585 & 2 & 17 & 1 & 241.652 & 51.3070 & 1.54E+10 & 9.67E+09 \\
152 & 4 & 2 & 4 & 241.740 & 51.2883 & 1.53E+10 & 9.45E+09 \\
252 & 3 & 8 & 2 & 242.348 & 51.1597 & 2.84E+10 & 1.75E+10 \\
220 & 1 & 6 & 0 & 242.624 & 51.1014 & 1.81E+10 & 2.02E+10 \\
149 & 4 & 3 & 3 & 242.874 & 51.0488 & 2.24E+10 & 1.41E+10 \\
220 & 1 & 7 & 2 & 242.891 & 51.0452 & 1.51E+10 & 8.58E+09 \\
288 & 4 & 9 & 4 & 242.963 & 51.0301 & 3.17E+10 & 2.00E+10 \\
645 & 3 & 19 & 2 & 243.258 & 50.9682 & 3.94E+10 & 2.54E+10 \\
1189 & 1 & 30 & 0 & 243.416 & 50.9352 & 1.02E+10 & 1.15E+10 \\
146 & 5 & 2 & 4 & 243.446 & 50.9289 & 1.31E+10 & 8.15E+09 \\
677 & 2 & 21 & 3 & 243.506 & 50.9163 & 1.57E+10 & 1.02E+10 \\
210 & 2 & 7 & 2 & 243.801 & 50.8547 & 1.38E+10 & 9.04E+09 \\
198 & 4 & 5 & 3 & 244.073 & 50.7980 & 2.43E+10 & 1.51E+10 \\
674 & 5 & 22 & 4 & 245.143 & 50.5762 & 1.19E+10 & 8.02E+09 \\
151 & 6 & 4 & 5 & 245.161 & 50.5727 & 3.74E+10 & 2.32E+10 \\
137 & 2 & 3 & 3 & 245.336 & 50.5365 & 1.39E+10 & 8.56E+09 \\
740 & 1 & 28 & 2 & 245.630 & 50.4760 & 1.93E+10 & 1.30E+10 \\
269 & 1 & 10 & 1 & 245.656 & 50.4707 & 3.05E+10 & 1.89E+10 \\
268 & 2 & 10 & 1 & 245.657 & 50.4703 & 2.09E+10 & 1.33E+10 \\
256 & 5 & 9 & 4 & 245.659 & 50.4701 & 3.46E+10 & 2.16E+10 \\
196 & 2 & 7 & 2 & 245.830 & 50.4349 & 2.43E+10 & 1.57E+10 \\
115 & 1 & 1 & 2 & 246.293 & 50.3401 & 2.73E+10 & 3.14E+10 \\
190 & 3 & 7 & 2 & 246.435 & 50.3111 & 2.38E+10 & 1.57E+10 \\
128 & 5 & 2 & 4 & 246.794 & 50.2380 & 1.49E+10 & 9.43E+09 \\
129 & 2 & 3 & 3 & 247.082 & 50.1794 & 1.18E+10 & 7.57E+09 \\
617 & 2 & 19 & 2 & 247.119 & 50.1718 & 1.60E+10 & 1.85E+10 \\
203 & 1 & 8 & 2 & 247.550 & 50.0846 & 2.18E+10 & 1.40E+10 \\
244 & 0 & 10 & 1 & 248.394 & 49.9144 & 3.30E+10 & 2.07E+10 \\
112 & 2 & 1 & 2 & 249.171 & 49.7587 & 1.40E+10 & 1.65E+10 \\
162 & 1 & 7 & 2 & 250.565 & 49.4819 & 1.17E+10 & 7.60E+09 \\
155 & 3 & 5 & 3 & 250.671 & 49.4609 & 2.31E+10 & 1.65E+10 \\
499 & 1 & 17 & 1 & 251.234 & 49.3501 & 2.90E+10 & 2.28E+10 \\
585 & 2 & 19 & 2 & 251.307 & 49.3356 & 3.28E+10 & 2.41E+10 \\
102 & 3 & 1 & 2 & 251.557 & 49.2867 & 3.24E+10 & 2.12E+10 \\
574 & 1 & 19 & 2 & 252.623 & 49.0787 & 4.59E+10 & 3.85E+10 \\
114 & 3 & 2 & 4 & 252.630 & 49.0774 & 2.25E+10 & 1.73E+10 \\
114 & 3 & 3 & 3 & 253.273 & 48.9528 & 3.20E+10 & 2.31E+10 \\
138 & 2 & 5 & 3 & 254.083 & 48.7967 & 2.58E+10 & 1.96E+10 \\
559 & 3 & 19 & 2 & 254.431 & 48.7301 & 1.17E+10 & 7.77E+09 \\
111 & 4 & 2 & 4 & 255.376 & 48.5496 & 2.86E+10 & 2.08E+10 \\
444 & 2 & 17 & 1 & 256.039 & 48.4240 & 1.13E+10 & 9.61E+09 \\
529 & 1 & 19 & 2 & 257.852 & 48.0836 & 1.16E+11 & 7.48E+10 \\
427 & 0 & 17 & 1 & 257.884 & 48.0775 & 3.10E+10 & 2.47E+10 \\
91 & 1 & 1 & 2 & 257.921 & 48.0705 & 5.84E+10 & 4.48E+10 \\
138 & 2 & 8 & 2 & 258.472 & 47.9682 & 3.35E+10 & 2.52E+10 \\
111 & 4 & 4 & 5 & 258.898 & 47.8892 & 5.35E+10 & 4.13E+10 \\
155 & 3 & 9 & 4 & 259.384 & 47.7996 & 5.15E+10 & 4.02E+10 \\
118 & 1 & 6 & 0 & 262.224 & 47.2818 & 2.63E+10 & 1.99E+10 \\
114 & 3 & 7 & 2 & 264.250 & 46.9193 & 2.37E+10 & 1.88E+10 \\
138 & 2 & 10 & 1 & 264.588 & 46.8593 & 2.46E+10 & 1.98E+10 \\
96 & 4 & 4 & 5 & 264.727 & 46.8347 & 1.15E+10 & 9.24E+09 \\
61 & 2 & 1 & 2 & 268.437 & 46.1875 & 1.77E+10 & 1.36E+10 \\
91 & 1 & 6 & 0 & 276.761 & 44.7982 & 2.25E+10 & 1.84E+10 \\
246 & 0 & 17 & 1 & 279.774 & 44.3159 & 7.08E+10 & 6.14E+10 \\
36 & 2 & 1 & 2 & 283.372 & 43.7531 & 3.88E+10 & 3.33E+10 \\
138 & 2 & 26 & 3 & 339.767 & 36.4910 & 1.59E+10 & 1.15E+10 \\
114 & 3 & 22 & 4 & 342.767 & 36.1715 & 1.40E+10 & 1.04E+10 \\
155 & 3 & 28 & 2 & 343.893 & 36.0531 & 1.08E+10 & 8.24E+09 \\
111 & 4 & 21 & 3 & 344.722 & 35.9665 & 1.17E+10 & 8.83E+09 \\
529 & 1 & 30 & 0 & 346.804 & 35.7505 & 1.13E+10 & 8.46E+09 \\
\end{longtable}

\clearpage

\setlength{\LTcapwidth}{\linewidth}
\setlength{\LTleft}{0pt}
\setlength{\LTright}{0pt} 
\setlength{\tabcolsep}{0.5\tabcolsep}
\renewcommand{\arraystretch}{1.0}

\begin{longtable}{@{\extracolsep{\fill}} llll c  rrrrr}
\caption{High-intensity M1 and E2 transitions of \Wix.\label{tab:Wix_trf}}\\
\multicolumn{2}{c}{Upper level} & \multicolumn{2}{c}{Lower level} &&&&&&\\
\cmidrule{1-2}\cmidrule{3-4}
No. & $J$ & No. & $J$ & Type & $\lambda$ (\AA) & $E$ (eV) & $A_l$ (s$^{-1}$) & $A_v$ (s$^{-1}$) & $A_M$ (s$^{-1}$) \\
\midrule
\endfirsthead
\multicolumn{10}{l}{Table \ref{tab:Wix_trf} (continued)}\\
\midrule
\multicolumn{2}{c}{Upper level} & \multicolumn{2}{c}{Lower level} &&&&\\
\cmidrule{1-2}\cmidrule{3-4}
No. & $J$ & No. & $J$ & Type & $\lambda$ (\AA) & $E$ (eV) & $A_l$ (s$^{-1}$) & $A_v$ (s$^{-1}$) & $A_M$ (s$^{-1}$) \\
\midrule
\endhead
\midrule
\endfoot
\bottomrule
\endlastfoot
30 & 0 & 1 & 2 & E2 & 497.506 & 24.9212 & 1.46E+02 & 1.01E+02 &  \\
30 & 0 & 7 & 2 & E2 & 574.197 & 21.5926 & 1.97E+01 & 1.85E+01 &  \\
30 & 0 & 8 & 2 & E2 & 589.867 & 21.0190 & 2.28E+01 & 2.26E+00 &  \\
29 & 0 & 7 & 2 & E2 & 621.539 & 19.9479 & 4.67E+02 & 2.01E+02 &  \\
29 & 0 & 8 & 2 & E2 & 639.941 & 19.3743 & 2.40E+01 & 3.16E+01 &  \\
28 & 2 & 2 & 4 & E2 & 809.774 & 15.3110 & 4.20E+01 & 3.16E+01 &  \\
28 & 2 & 3 & 3 & E2 & 816.419 & 15.1863 & 3.60E+01 & 8.19E+00 &  \\
28 & 2 & 3 & 3 & M1 & 816.419 & 15.1863 &  &  & 1.24E+02 \\
27 & 2 & 2 & 4 & E2 & 857.572 & 14.4576 & 7.96E+00 & 6.43E+00 &  \\
27 & 2 & 3 & 3 & E2 & 865.028 & 14.3330 & 1.98E+00 & 3.47E+00 &  \\
30 & 0 & 17 & 1 & M1 & 866.754 & 14.3044 &  &  & 4.20E+04 \\
26 & 3 & 2 & 4 & E2 & 872.578 & 14.2090 & 1.10E+01 & 8.31E+00 &  \\
26 & 3 & 2 & 4 & M1 & 872.578 & 14.2090 &  &  & 4.25E+02 \\
26 & 3 & 3 & 3 & E2 & 880.298 & 14.0843 & 9.28E+00 & 8.86E+00 &  \\
26 & 3 & 3 & 3 & M1 & 880.298 & 14.0843 &  &  & 1.04E+02 \\
25 & 1 & 3 & 3 & E2 & 889.535 & 13.9381 & 1.54E+01 & 2.16E+01 &  \\
21 & 3 & 1 & 2 & E2 & 901.470 & 13.7536 & 2.39E+00 & 4.21E-01 &  \\
28 & 2 & 5 & 3 & E2 & 924.722 & 13.4077 & 3.34E+02 & 2.60E+02 &  \\
28 & 2 & 5 & 3 & M1 & 924.722 & 13.4077 &  &  & 5.36E+03 \\
23 & 6 & 2 & 4 & E2 & 938.189 & 13.2153 & 9.99E+00 & 1.33E+01 &  \\
28 & 2 & 7 & 2 & E2 & 942.643 & 13.1528 & 9.79E+00 & 8.98E+00 &  \\
28 & 2 & 7 & 2 & M1 & 942.643 & 13.1528 &  &  & 6.07E+02 \\
22 & 4 & 2 & 4 & E2 & 960.681 & 12.9059 & 2.57E+02 & 2.30E+02 &  \\
22 & 4 & 2 & 4 & M1 & 960.681 & 12.9059 &  &  & 4.72E+03 \\
29 & 0 & 16 & 2 & E2 & 961.042 & 12.9010 & 1.88E+01 & 2.24E-02 &  \\
22 & 4 & 3 & 3 & E2 & 970.048 & 12.7812 & 3.01E+02 & 2.33E+02 &  \\
22 & 4 & 3 & 3 & M1 & 970.048 & 12.7812 &  &  & 1.57E+03 \\
19 & 2 & 1 & 2 & E2 & 984.933 & 12.5881 & 5.79E+02 & 5.54E+02 &  \\
19 & 2 & 1 & 2 & M1 & 984.933 & 12.5881 &  &  & 7.56E+03 \\
21 & 3 & 2 & 4 & E2 & 985.321 & 12.5831 & 2.66E+02 & 2.19E+02 &  \\
21 & 3 & 2 & 4 & M1 & 985.321 & 12.5831 &  &  & 4.03E+03 \\
28 & 2 & 8 & 2 & E2 & 985.627 & 12.5792 & 5.44E+01 & 7.22E+01 &  \\
28 & 2 & 8 & 2 & M1 & 985.627 & 12.5792 &  &  & 5.85E+03 \\
27 & 2 & 5 & 3 & E2 & 987.580 & 12.5543 & 2.12E+01 & 2.38E+01 &  \\
21 & 3 & 3 & 3 & E2 & 995.177 & 12.4585 & 8.19E+01 & 9.43E+01 &  \\
21 & 3 & 3 & 3 & M1 & 995.177 & 12.4585 &  &  & 5.95E+03 \\
30 & 0 & 19 & 2 & E2 & 1005.297 & 12.3331 & 2.25E+03 & 1.14E+03 &  \\
26 & 3 & 5 & 3 & E2 & 1007.533 & 12.3057 & 2.46E+02 & 2.20E+02 &  \\
26 & 3 & 5 & 3 & M1 & 1007.533 & 12.3057 &  &  & 4.20E+03 \\
27 & 2 & 7 & 2 & E2 & 1008.047 & 12.2995 & 1.58E+00 & 4.05E+00 &  \\
22 & 4 & 4 & 5 & E2 & 1012.493 & 12.2454 & 1.24E+02 & 1.34E+02 &  \\
22 & 4 & 4 & 5 & M1 & 1012.493 & 12.2454 &  &  & 9.28E+03 \\
20 & 2 & 2 & 4 & E2 & 1027.135 & 12.0709 & 5.02E+00 & 5.07E+00 &  \\
26 & 3 & 7 & 2 & E2 & 1028.844 & 12.0508 & 1.67E+01 & 1.84E+01 &  \\
26 & 3 & 7 & 2 & M1 & 1028.844 & 12.0508 &  &  & 9.64E+01 \\
20 & 2 & 3 & 3 & E2 & 1037.849 & 11.9463 & 2.50E+01 & 2.73E+01 &  \\
21 & 3 & 4 & 5 & E2 & 1039.900 & 11.9227 & 4.56E+02 & 3.74E+02 &  \\
28 & 2 & 9 & 4 & E2 & 1055.506 & 11.7464 & 4.44E+02 & 3.75E+02 &  \\
27 & 2 & 8 & 2 & E2 & 1057.359 & 11.7258 & 1.17E+01 & 1.48E+01 &  \\
24 & 0 & 7 & 2 & E2 & 1069.044 & 11.5977 & 5.72E+01 & 4.77E+01 &  \\
26 & 3 & 8 & 2 & E2 & 1080.264 & 11.4772 & 1.60E+02 & 1.56E+02 &  \\
26 & 3 & 8 & 2 & M1 & 1080.264 & 11.4772 &  &  & 1.04E+03 \\
28 & 2 & 10 & 1 & E2 & 1080.911 & 11.4703 & 6.31E+00 & 1.19E+01 &  \\
28 & 2 & 10 & 1 & M1 & 1080.911 & 11.4703 &  &  & 4.58E+03 \\
19 & 2 & 2 & 4 & E2 & 1085.899 & 11.4177 & 1.42E+00 & 4.55E-01 &  \\
25 & 1 & 8 & 2 & E2 & 1094.207 & 11.3310 & 9.48E+00 & 1.01E+01 &  \\
19 & 2 & 3 & 3 & E2 & 1097.882 & 11.2930 & 3.85E+00 & 1.21E-01 &  \\
24 & 0 & 8 & 2 & E2 & 1124.669 & 11.0241 & 1.59E+01 & 2.39E+01 &  \\
22 & 4 & 5 & 3 & E2 & 1126.860 & 11.0026 & 1.30E+00 & 3.36E+00 &  \\
22 & 4 & 5 & 3 & M1 & 1126.860 & 11.0026 &  &  & 4.46E+01 \\
22 & 4 & 7 & 2 & E2 & 1153.585 & 10.7477 & 1.15E+02 & 1.14E+02 &  \\
21 & 3 & 5 & 3 & M1 & 1160.913 & 10.6799 &  &  & 8.97E+01 \\
26 & 3 & 9 & 4 & E2 & 1164.781 & 10.6444 & 8.28E+01 & 1.01E+02 &  \\
26 & 3 & 9 & 4 & M1 & 1164.781 & 10.6444 &  &  & 8.66E+03 \\
17 & 1 & 1 & 2 & E2 & 1167.817 & 10.6168 & 2.50E+02 & 2.35E+02 &  \\
17 & 1 & 1 & 2 & M1 & 1167.817 & 10.6168 &  &  & 1.05E+04 \\
18 & 4 & 3 & 3 & E2 & 1180.418 & 10.5034 & 1.56E+00 & 1.45E+00 &  \\
21 & 3 & 7 & 2 & E2 & 1189.297 & 10.4250 & 6.25E+00 & 1.47E+01 &  \\
21 & 3 & 7 & 2 & M1 & 1189.297 & 10.4250 &  &  & 4.21E+03 \\
26 & 3 & 10 & 1 & E2 & 1195.796 & 10.3683 & 7.96E+01 & 9.02E+01 &  \\
25 & 1 & 10 & 1 & E2 & 1212.905 & 10.2221 & 2.08E+01 & 2.58E+01 &  \\
22 & 4 & 8 & 2 & E2 & 1218.623 & 10.1741 & 7.58E+00 & 9.57E+00 &  \\
29 & 0 & 20 & 2 & E2 & 1235.496 & 10.0352 & 1.97E+01 & 1.01E-02 &  \\
20 & 2 & 7 & 2 & E2 & 1250.755 & 9.9127 & 3.89E+00 & 5.80E+00 &  \\
21 & 3 & 8 & 2 & M1 & 1258.546 & 9.8514 &  &  & 4.14E+02 \\
23 & 6 & 9 & 4 & E2 & 1284.714 & 9.6507 & 1.87E+01 & 3.00E+01 &  \\
19 & 2 & 5 & 3 & E2 & 1303.119 & 9.5144 & 1.11E+01 & 2.00E+00 &  \\
22 & 4 & 9 & 4 & E2 & 1327.266 & 9.3413 & 3.32E+00 & 3.37E+00 &  \\
22 & 4 & 9 & 4 & M1 & 1327.266 & 9.3413 &  &  & 1.30E+02 \\
20 & 2 & 8 & 2 & E2 & 1327.577 & 9.3391 & 3.36E+00 & 3.50E+00 &  \\
19 & 2 & 6 & 0 & E2 & 1330.910 & 9.3157 & 3.85E+01 & 9.19E+01 &  \\
15 & 3 & 2 & 4 & E2 & 1352.803 & 9.1650 & 1.51E+00 & 3.05E+00 &  \\
16 & 2 & 3 & 3 & E2 & 1365.401 & 9.0804 & 3.83E+00 & 4.61E+00 &  \\
21 & 3 & 9 & 4 & E2 & 1374.763 & 9.0186 & 2.65E+00 & 2.43E+00 &  \\
21 & 3 & 9 & 4 & M1 & 1374.763 & 9.0186 &  &  & 1.43E+02 \\
14 & 4 & 2 & 4 & E2 & 1419.548 & 8.7341 & 1.55E+00 & 2.89E+00 &  \\
18 & 4 & 5 & 3 & E2 & 1421.056 & 8.7248 & 1.80E+00 & 3.24E+00 &  \\
19 & 2 & 8 & 2 & E2 & 1427.417 & 8.6859 & 1.79E+00 & 7.97E-01 &  \\
30 & 0 & 28 & 2 & E2 & 1469.045 & 8.4398 & 5.14E+01 & 7.37E+00 &  \\
20 & 2 & 10 & 1 & E2 & 1506.443 & 8.2303 & 2.95E+00 & 6.65E+00 &  \\
29 & 0 & 25 & 1 & M1 & 1541.450 & 8.0434 &  &  & 1.13E+03 \\
19 & 2 & 9 & 4 & E2 & 1578.789 & 7.8531 & 9.70E+00 & 1.93E+00 &  \\
29 & 0 & 27 & 2 & E2 & 1621.031 & 7.6485 & 8.63E+00 & 6.85E-03 &  \\
17 & 1 & 5 & 3 & E2 & 1643.681 & 7.5431 & 5.34E+00 & 6.12E-01 &  \\
23 & 6 & 11 & 6 & M1 & 1678.252 & 7.3877 &  &  & 7.10E+01 \\
17 & 1 & 6 & 0 & M1 & 1688.144 & 7.3444 &  &  & 1.77E+03 \\
15 & 3 & 5 & 3 & E2 & 1707.361 & 7.2617 & 1.31E+00 & 2.84E+00 &  \\
18 & 4 & 9 & 4 & E2 & 1755.282 & 7.0635 & 1.04E+00 & 3.98E+00 &  \\
16 & 2 & 7 & 2 & E2 & 1759.411 & 7.0469 & 4.48E+00 & 1.16E+01 &  \\
12 & 4 & 2 & 4 & E2 & 1782.104 & 6.9572 & 2.34E+00 & 6.30E+00 &  \\
12 & 4 & 3 & 3 & E2 & 1814.608 & 6.8326 & 3.54E+00 & 9.88E+00 &  \\
14 & 4 & 5 & 3 & E2 & 1815.070 & 6.8308 & 1.12E+00 & 3.32E+00 &  \\
29 & 0 & 28 & 2 & E2 & 1824.612 & 6.7951 & 4.37E+00 & 8.03E-01 &  \\
17 & 1 & 8 & 2 & E2 & 1846.493 & 6.7146 & 3.14E+00 & 5.72E-01 &  \\
15 & 3 & 8 & 2 & E2 & 1927.243 & 6.4332 & 1.62E+00 & 5.35E+00 &  \\
13 & 5 & 5 & 3 & E2 & 2100.027 & 5.9039 & 1.23E+00 & 4.51E+00 &  \\
11 & 6 & 2 & 4 & E2 & 2127.546 & 5.8276 & 1.63E+00 & 5.86E+00 &  \\
17 & 1 & 10 & 1 & E2 & 2211.751 & 5.6057 & 1.00E+00 & 3.58E-01 &  \\
23 & 6 & 13 & 5 & M1 & 2292.568 & 5.4081 &  &  & 2.04E+01 \\
27 & 2 & 15 & 3 & M1 & 2342.597 & 5.2926 &  &  & 1.32E+02 \\
25 & 1 & 16 & 2 & M1 & 2552.343 & 4.8577 &  &  & 3.53E+01 \\
18 & 4 & 12 & 4 & M1 & 3377.533 & 3.6709 &  &  & 7.06E+00 \\
9 & 4 & 3 & 3 & M1 & 3604.276 & 3.4399 &  &  & 1.20E+01 \\
6 & 0 & 1 & 2 & E2 & 3788.855 & 3.2723 & 2.39E+00 & 2.14E-01 &  \\
20 & 2 & 15 & 3 & M1 & 4266.645 & 2.9059 &  &  & 1.08E+02 \\
9 & 4 & 4 & 5 & M1 & 4269.259 & 2.9041 &  &  & 1.79E+02 \\
20 & 2 & 16 & 2 & M1 & 4326.289 & 2.8658 &  &  & 7.02E+01 \\
18 & 4 & 13 & 5 & M1 & 4395.268 & 2.8209 &  &  & 1.59E+02 \\
28 & 2 & 21 & 3 & M1 & 4545.155 & 2.7278 &  &  & 1.42E+02 \\
8 & 2 & 3 & 3 & M1 & 4755.593 & 2.6071 &  &  & 9.60E+01 \\
27 & 2 & 20 & 2 & M1 & 5194.792 & 2.3867 &  &  & 8.65E+01 \\
15 & 3 & 12 & 4 & M1 & 5615.719 & 2.2078 &  &  & 8.48E+01 \\
25 & 1 & 20 & 2 & M1 & 6224.638 & 1.9918 &  &  & 3.66E+01 \\
13 & 5 & 11 & 6 & M1 & 6263.078 & 1.9796 &  &  & 1.07E+02 \\
19 & 2 & 17 & 1 & M1 & 6289.352 & 1.9713 &  &  & 1.31E+01 \\
5 & 3 & 2 & 4 & M1 & 6514.374 & 1.9032 &  &  & 6.84E+01 \\
18 & 4 & 14 & 4 & M1 & 6546.268 & 1.8940 &  &  & 3.54E+01 \\
5 & 3 & 3 & 3 & M1 & 6970.808 & 1.7786 &  &  & 6.88E+00 \\
14 & 4 & 12 & 4 & M1 & 6977.622 & 1.7769 &  &  & 3.48E+01 \\
10 & 1 & 7 & 2 & M1 & 7369.094 & 1.6825 &  &  & 1.94E+01 \\
9 & 4 & 5 & 3 & M1 & 7463.088 & 1.6613 &  &  & 5.43E+00 \\
26 & 3 & 21 & 3 & M1 & 7625.898 & 1.6258 &  &  & 2.10E+00 \\
18 & 4 & 15 & 3 & M1 & 8474.397 & 1.4630 &  &  & 3.11E+00 \\
26 & 3 & 22 & 4 & M1 & 9514.603 & 1.3031 &  &  & 1.52E+01 \\
10 & 1 & 8 & 2 & M1 & 11181.055 & 1.1089 &  &  & 1.64E+01 \\
14 & 4 & 13 & 5 & M1 & 13376.387 & 0.9269 &  &  & 6.84E+00 \\
8 & 2 & 5 & 3 & M1 & 14964.818 & 0.8285 &  &  & 4.94E+00 \\
\end{longtable}

\end{document}